\documentclass[12pt,a4paper]{article}

\usepackage{jheppub}

\usepackage{etoolbox}
    \makeatletter
    \patchcmd{\maketitle}{\@fpheader}{}{}{}
    \makeatother

\usepackage[utf8]{inputenc}
\usepackage{lmodern}
\usepackage{exscale}

\usepackage{amsmath}
\usepackage{amssymb}
\usepackage{mathtools}

\usepackage{mathrsfs}

\usepackage{tensor}

\usepackage[low-sup]{subdepth}

\newcommand{\scr}{\scriptscriptstyle}

\usepackage{graphicx}

\usepackage{xcolor}
\hypersetup{
    colorlinks,
    linkcolor={black},
    citecolor={black},
    urlcolor={black}
}

\usepackage[normalem]{ulem}

\usepackage{cancel}

\makeatletter
\newcommand{\dalembertian}{\mathop{\mathpalette\dalembertian@\relax}}
\newcommand{\dalembertian@}[2]{%
  \begingroup
  \sbox\z@{$\m@th#1\square$}%
  \dimen0=\fontdimen8
    \ifx#1\displaystyle\textfont\else
    \ifx#1\textstyle\textfont\else
    \ifx#1\scriptstyle\scriptfont\else
    \scriptscriptfont\fi\fi\fi3
  \makebox[\wd\z@]{%
    \hbox to \ht\z@{%
      \vrule width \dimen0
      \kern-\dimen0
      \vbox to \ht\z@{
        \hrule height \dimen0 width \ht\z@
        \vss
        \hrule height 2\dimen0
      }%
      \kern-2.5\dimen0
      \vrule width 2.5\dimen0
    }%
  }%
  \endgroup
}
\makeatother

\begin{document}

\title{Photon quantization in cosmological spaces}

\author[]{Dra\v{z}en Glavan,}
\emailAdd{glavan@fzu.cz}


\affiliation[]{CEICO, FZU --- Institute of Physics of the Czech Academy of Sciences,
	\\
	Na Slovance 1999/2, 182 21 Prague 8, Czech Republic}

\abstract{
Canonical quantization of the photon --- a free massless vector field --- is considered in 
cosmological spacetimes in a two-parameter family of linear gauges 
that treat all the 
vector potential components on equal footing. The goal is setting up a
framework for computing photon two-point functions appropriate for loop computations
in realistic inflationary spacetimes.
The quantization is implemented without relying on spacetime symmetries, 
but rather it is based on the classical canonical structure.
Special attention is paid to the quantization of the canonical first-class constraint structure
that is implemented as the condition on the physical states.
This condition gives rise to subsidiary conditions 
that the photon two-point functions must satisfy.
Some of the de Sitter space photon propagators from the literature are found not to satisfy 
these subsidiary conditions, bringing into question their consistency.
}

\notoc
\maketitle
\titlepage

\section{Introduction}
\label{sec: Introduction}

The massless vector field --- the photon ---  couples conformally to gravity in four spacetime 
dimensions, and effectively does not see the cosmological expansion. 
That is why its linear dynamics is not as interesting 
compared to nonconformally coupled fields
that experience gravitational particle 
production~\cite{Parker:1968mv,Parker:1969au,Parker:1971pt}
at the linear level. 
Nevertheless, coupling of photons to
other fields might break conformality.
Particularly interesting cases occur in inflation where the conformal coupling 
of the photon
is broken by quantum loops generated
from its interactions with light spectator scalars~\cite{Turner:1987bw,Calzetta:1997ku,Kandus:1999st,Giovannini:2000dj,Dimopoulos:2001wx,Davis:2000zp,Maroto:2000zu,Dolgov:2001ce,Calzetta:2001cf,Prokopec:2002jn,Prokopec:2002uw,Prokopec:2003bx,Prokopec:2003iu,Kahya:2005kj,Kahya:2006ui,Prokopec:2006ue,Prokopec:2007ak,Prokopec:2008gw,Leonard:2012si,Leonard:2012ex,Chen:2016nrs,Chen:2016uwp,Chen:2016hrz,Kaya:2018qbj,Popov:2017xut,Glavan:2019uni}
or gravitons~\cite{Leonard:2013xsa,Glavan:2013jca,Wang:2014tza,Glavan:2015ura,Glavan:2016bvp,Miao:2018bol,Glavan:2023tet}. 
Scalars and gravitons are nonconformally coupled to gravity 
and experience a huge enhancement of their infrared sector due to the rapid expansion during inflation,
the effects of which are communicated to the photon via loops. 
Most of the
loop computations 
involving photons
thus far have been worked out
in a rigid de Sitter background, for which the two-point functions (propagators) 
comprising the loop expansion are known.
It would be interesting to understand how the results,
in particular the secular corrections to photons,
are modulated in realistic
slow-roll inflationary spacetimes. This work aims to advance this understanding by considering
the photon and its two-point functions in a two-parameter family of linear 
gauges in general expanding cosmological spaces.

The specific goals of this article are threefold:
\begin{itemize}
\item[(i)]
To understand the photon quantization without relying on background symmetries or 
on covariant gauge fixing;

\item[(ii)]
To set up a framework that will facilitate the computations of photon two-point functions in 
$D$-dimensional cosmological spaces, and to derive all the subsidiary conditions the two-point functions 
must satisfy, with the goal of performing dimensionally regulated loop computations in realistic inflationary
spacetimes;

\item[(iii)]
To demonstrate that cosmological evolution is not in conflict with gauge invariance,
as suggested in~\cite{BeltranJimenez:2008enx},
and that there is no contribution to the photon energy-momentum tensor coming from the 
gauge-fixing terms in any cosmological spacetime.

\end{itemize}
Given the aims listed above, 
in this work we consider the photon (electromagnetism) is spatially flat cosmological spacetimes in 
a two-parameter family of linear gauges preserving cosmological symmetries,

\begin{equation}
S_{\rm gf}[A_\mu] = \int\! d^{D\!}x \, \sqrt{-g} \, \biggl[
	- \frac{1}{2\xi} \Bigl( \nabla^\mu A_\mu - 2 \zeta n^\mu A_\mu \Bigr)^{\!2 \,}
	\biggr] 
	\, ,
\label{gauge-fixing term}
\end{equation}
where~$\xi$ and~$\zeta$ are arbitrary gauge-fixing parameters, and where~$n_\mu\!=\!\delta_\mu^0 \mathcal{H}$ is
a nondynamical timelike vector, invariant under spatial rotations and translations.
In the limit~$\zeta\!=\!0$ this gauge-fixing functional reduces to the general covariant gauge,
which one might be tempted to consider as a natural choice.
However, experience in de Sitter space suggests that noncovariant gauge choices
can lead to simpler two-point functions~\cite{Woodard:2004ut} that can considerably simplify
often quite involved computations. To this end we set up the framework that will facilitate identifying simple gauges
in~$D$-dimensional inflationary spacetimes, and streamline their computation.

In de Sitter space both the photon and the massive vector two-point functions 
have been worked out: in general covariant gauge~\cite{Allen:1985wd,Tsamis:2006gj,Youssef:2010dw,Frob:2013qsa}, 
and in the simple gauge~\cite{Woodard:2004ut},
while for more general spacetimes only a few results exist --- the unitary gauge massive vector
propagator~\cite{Glavan:2020zne} in power-law inflation, and photon propagator in arbitrary
4-dimensional Friedmann-Lama\^{i}tre-Robertson-Walker (FLRW) spacetime in
a conformal gauge~\cite{Huguet:2013dia} that takes the same form as in flat space on the account of 
conformal coupling. Exact gauges, such as the Coulomb one in which the two-point function takes the same form 
as in flat space in four-dimensional spacetime~\cite{Cotaescu:2008hv}, are also legitimate choices.
However, for practical applications the~$D$-dimensional multiplier gauges such as~(\ref{gauge-fixing term})
are usually preferable.

Because of the noncovariant gauge-fixing~(\ref{gauge-fixing term}), and because 
of the reduced symmetry of cosmological spaces compared to maximally symmetric ones, 
we have to understand how to compute photon two-point functions without relying on a large number
of symmetries that usually simplify problems. To this end we consider the canonical quantization 
of the photon field similar to the Gupta-Bleuler quantization~\cite{Gupta:1949rh,Bleuler:1950cy}.
The approach taken here mainly differs by not relying on the on symmetries of the system.
Rather it is based on the classical canonical structure in the {\it multiplier gauge} defined by
the gauge-fixing functional~(\ref{gauge-fixing term}).
In that sense the quantization outlined here can be considered as the derivation of the
Gupta-Bleuler quantization. This includes the derivation of the somewhat {\it ad hoc} Gupta-Bleuler
subsidiary condition on the physical space of states, 
that here arises as a natural consequence of the canonical first-class constraint structure 
and the correspondence principle. 

Several works in recent years have considered 
the quantization of the photon field in cosmological 
spaces~\cite{BeltranJimenez:2008enx,Frob:2013qsa,Finster:2013fva,Zhang:2022csl}, 
choosing to preserve general covariance as much as possible. While there is nothing wrong
with maintaining covariance, for practical purposes this is not always the most convenient choice,
and covariant approaches offer little help when abandoning manifest covariance by gauge choices.
This is why we opt to consider the canonical quantization formalism from the ground up,
divorced from spacetime symmetries.
We only consider spatially flat FLRW spacetimes, and not spatially closed ones where
the issue of linearization instability arises~\cite{Moncrief:1976un,Arms:1979au,Higuchi:1991tk,Higuchi:1991tm,Marolf:1995cn,Marolf:2008it,Marolf:2008hg,Miao:2009hb,Gibbons:2014zya,Altas:2021htf}.

Proper understanding of photon quantization has practical consequences. Computations
of quantum loop corrections in inflation are notoriously difficult, and the choice
of gauge can make a world of difference. It is sufficient to compare the 
computations of the one-graviton-loop vacuum polarization in de Sitter in
noncovariant gauges~\cite{Leonard:2013xsa} and covariant gauges~\cite{Glavan:2015ura}
to realize that de Sitter symmetries do not play the same convenient organizational role 
as do Poincar\'{e} symmetries in flat space. Therefore, it is advantageous to consider different
gauges in cosmological spacetimes. We ultimately express
the photon two-point functions in terms of a few scalar mode functions that cosmologists 
are accustomed to working with. Identifying gauges admitting simple solutions for these 
scalar mode functions is left for future work.

The two-point functions of the linear theory serve to compute loop corrections to physical observables.
It is in no way obvious for gauge theories in multiplier gauges, such 
as~(\ref{gauge-fixing term}), how to properly define quantum observables. It was suggested
in~\cite{BeltranJimenez:2008enx} that the expectation value of the photon 
energy-momentum tensor 
depends on whether the contribution from the gauge-fixing term~(\ref{gauge-fixing term}) is
included or not. For covariant gauges (with~$\zeta\!=\!0$) in de Sitter, the gauge-fixing term is 
supposed to contribute as a cosmological constant to the energy-momentum tensor. 
This is in conflict with the correspondence principle, and would essentially allow 
for quantum measurements of first-class
constraints. However, when ordering of products of field operators comprising the energy-momentum
tensor is considered carefully, as in Sec.~\ref{sec: Simple observables}, the issue is fully resolved. The gauge-fixing
functional~(\ref{gauge-fixing term}) cannot contribute to the energy-momentum tensor for any admissible state,
in any cosmological spacetime.

The results of this work also include the unexpected observation that the subsidiary conditions for the 
photon two-point functions derived here are not satisfied for some of the de Sitter space
photon propagators reported in the literature. More details on this issue are given by the end of the
concluding section~\ref{sec: Discussion}, while further investigations into the issue are left for future 
work~\cite{Glavan:2022dwb,Glavan:2022nrd}.


The paper is organized in nine sections, with the current one laying out the background and
the motivation.
Section~\ref{sec: Preliminaries}
gives some of the properties of scalar field modes and two-point 
functions that we make use of in subsequent sections. 
Section~\ref{sec: Classical photon in FLRW}
presents the details of implementing multiplier/average gauges in the canonical 
formulation of the classical photon. This structure lends itself to canonical quantization
that is discussed in Sec.~\ref{sec: Quantized photon in FLRW}.
In Sec.~\ref{sec: Field operator dynamics} the dynamics of the field operators
is translated into equations of motion for the mode functions, while 
Sec.~\ref{sec: Constructing space of states}
discusses the construction of the space of states and the conditions the
physical states must respect, as well as 
the role that spacetime symmetries play in this construction.
Section~\ref{sec: Two-point functions}
concerns the main goal of this paper of constructing photon two-point
functions from the photon mode functions, and demonstrates how such construction 
satisfies all the subsidiary conditions. 
Section~\ref{sec: Simple observables}
discusses two simple observables 
and the issue of proper operator ordering of observables. 
Section~\ref{sec: Discussion}
contains the discussion of the construction presented in the paper, and provides the check of 
the photon propagators in the literature versus the conditions presented here, not all of which are found to be satisfied.

\section{Preliminaries}
\label{sec: Preliminaries}

The mode functions of linear higher spin fields in cosmological spaces can often be expressed
in terms of the scalar mode functions. Consequently, the same is often true for two-point
functions as well. This indeed is the case for the photon mode functions and two-point functions
that we consider in this work.
This section serves to introduce the background cosmological space,
to define the notation, and to summarize some of the basic results for scalar fields
that will be used in subsequent sections.

\subsection{FLRW spacetime}
\label{subsec: FLRW spacetime}

The geometry of homogeneous, isotropic, and spatially flat expanding spacetimes
is described by the FLRW invariant line element,
\begin{equation}
ds^2
	= - dt^2 + a^2(t) d\vec{x}^{\,2} 
	= a^2(\eta) \Bigl( - d\eta^2 + d\vec{x}^{\,2} \Bigr) \, .
\end{equation}
The flat spatial sections are covered by~$(D\!-\!1)$ Cartesian coordinates,~$x_i \!\in\! (-\infty, \infty)$,
while the evolution is expressed in terms of either the physical time~$t$ 
or the conformal time~$\eta$.
The dynamics of the expansion is encoded in the scale factor~$a$, that also provides the
connection between the two time coordinates,~$dt \!=\! a d\eta$. For our purposes the
conformal time coordinate is preferable, since then the FLRW 
metric,~$g_{\mu\nu}\!=\! a^2(\eta) \eta_{\mu\nu}$, is conformally flat, 
where~$\eta_{\mu\nu}\!=\!{\rm diag}(-1,1,\dots,1)$ is the~$D$-dimensional Minkowski metric.

The relevant information about the dynamics of the expansion is usually captured by the first few
derivatives of the scale factor. The conformal Hubble rate~$\mathcal{H}$
and the principal slow-roll parameter~$\epsilon$ capture the first two derivatives of the
scale factor,
\begin{equation}
\mathcal{H}(\eta) = \frac{1}{a} \frac{d a}{d \eta} \, ,
\qquad \qquad
\epsilon(\eta) = 1 - \frac{1}{\mathcal{H}^2} \frac{d \mathcal{H}}{d\eta} \, .
\end{equation}
The more commonly used physical Hubble rate~$H$ is related to the conformal Hubble rate as~$H \!=\! \mathcal{H}/a$,
while the principal slow-roll parameter is related to the often used deceleration 
parameter,~$q\!=\! \epsilon \!-\! 1$. Accelerating FLRW spacetimes, such as primordial inflation,
are characterized by~$0\!<\!\epsilon(\eta)\!\ll\! 1$. 
Even though the results of this
work apply to arbitrary accelerating FLRW spacetimes, it is nonetheless helpful at
times to have a concrete spacetime in mind. One such tractable example is that
of power-law inflation~\cite{Lucchin:1984yf,La:1989za}
for which the Hubble rate and the scale factor depend on time as
\begin{equation}
\epsilon = {\tt const}
\qquad \Longrightarrow \qquad
\mathcal{H}(\eta) = \frac{H_0}{1 \!-\! (1\!-\!\epsilon)H_0(\eta \!-\! \eta_0)} \, ,
\qquad
a(\eta) = \Bigl( \frac{\mathcal{H}}{H_0} \Bigr)^{\frac{1}{1-\epsilon}}
\, ,
\label{power-law quants}
\end{equation}
where~$\eta_0$ is the initial time at which~$a(\eta_0)\!=\!1$ 
and~$\mathcal{H}(\eta_0)\!=\!H_0$.

\subsection{Scalar mode functions}
\label{subsec: Scalar mode functions}

The equation of motion for the conformally  rescaled scalar field modes 
of comoving momentum~$\vec{k}$ in FLRW generically takes the form
\begin{equation}
\biggl[
	\partial_0^2 + k^2 - \Bigl( \lambda^2 \!-\! \frac{1}{4} \Bigr) (1\!-\!\epsilon)^2 \mathcal{H}^2
	\biggr] \mathscr{U}_\lambda(\eta,\vec{k})
	= 0 \, ,
	\qquad \quad
	k = \| \vec{k} \| \, ,
\label{scalar mode eq}
\end{equation}
where~$\lambda$ descends from the mass term and the nonminimal coupling term.
Both~$\epsilon$ and~$\lambda$ are time-dependent in general, though
there are notable cases where they are constant:
in power-law inflation~$\epsilon$ is constant, while 
a massless, nonminimally coupled scalar has a constant $\lambda$.
The equation of motion~(\ref{scalar mode eq}) admits two linearly independent solutions,
which we take to be complex conjugates of each other,
\begin{equation}
\mathscr{U}_\lambda(\eta,\vec{k}) = \alpha(\vec{k}) \, \mathcal{U}_\lambda(\eta,k)
	+ \beta(\vec{k}) \, \mathcal{U}_\lambda^*(\eta,k) \, ,
\label{general solution}
\end{equation}
where~$\alpha(\vec{k})$ and~$\beta(\vec{k})$ are arbitrary constants of integration.
In practice it is convenient to choose the independent solutions,~$\mathcal{U}_\lambda$
and~$\mathcal{U}_\lambda^*$,
that are some appropriate generalization of positive- and negative-frequency
modes (Chernikov-Tagirov-Bunch-Davies modes~\cite{Chernikov:1968zm,Bunch:1978yq}), at least in the ultraviolet. 
As a concrete example one can keep in mind power-law
inflation~(\ref{power-law quants})
where~$0\!<\!\epsilon\!=\!{\tt const}\!<\!1$,
for which the positive-frequency mode function is~\footnote{The phase factor in the
definition~(\ref{U power-law def}) ensures that the Wronskian in~(\ref{Wronskian or})
is correct for both real and imaginary values of~$\lambda$.}
\begin{equation}
\epsilon = {\tt const}
\qquad \Longrightarrow \qquad
\mathcal{U}_{\lambda}(\eta,k) = 
	e^{\frac{i\pi}{4} (2\lambda+1) }
	\sqrt{ \frac{\pi}{4(1\!-\!\epsilon) \mathcal{H}} } \,
	H_\lambda^{\scr (1)} \biggl( \frac{k}{(1\!-\!\epsilon) \mathcal{H}} \biggr) \, ,
\label{U power-law def}
\end{equation}
where~$H_\lambda^{\scr (1)}$ is the Hankel function of the first kind.
Note, however, that no assumptions on~$\epsilon$ are made throughout the paper.
Equation~(\ref{scalar mode eq}) also implies a nonvanishing Wronskian for the two 
independent solutions,
\begin{equation}
\mathcal{U}_\lambda(\eta,k) \partial_0 \mathcal{U}_\lambda^*(\eta,k)
	- \mathcal{U}_\lambda^*(\eta,k) \partial_0 \mathcal{U}_\lambda(\eta,k)
	= i
	\, ,
\label{Wronskian or}
\end{equation}
where the normalization is chosen for convenience, as appropriate for mode functions of
scalar field operators. 
The free coefficients in~(\ref{general solution}) then have an interpretation
of Bogolyubov coefficients that have to satisfy
\begin{equation}
|\alpha (\vec{k})|^2 - |\beta (\vec{k})|^2 = 1 \, .
\end{equation}

When solving for the photon field mode functions in Sec.~\ref{subsec: Solving for dynamics} we will 
encounter scalar mode equations~(\ref{scalar mode eq}) with only special combinations 
of parameters for which
\begin{equation}
\partial_0 \biggl[ \Bigl( \lambda \!+\! \frac{1}{2} \Bigr)(1\!-\!\epsilon) \biggr] = 0 \, .
\label{time relation}
\end{equation}
%
That allows us to introduce convenient recurrence relations for
contiguous mode functions without solving equations of motion explicitly,
\begin{equation}
\biggl[ \partial_0 + \Bigl( \lambda \!+\! \frac{1}{2} \Bigr)(1\!-\!\epsilon) \mathcal{H} \biggr]
	\mathcal{U}_\lambda
	\!=\! - i k \, \mathcal{U}_{\lambda+1} \, ,
\quad 
\biggl[ \partial_0 - \Bigl( \lambda \!+\! \frac{1}{2} \Bigr)(1\!-\!\epsilon) \mathcal{H} \biggr]
	\mathcal{U}_{\lambda+1}
	\!=\! - i k \, \mathcal{U}_{\lambda} \, ,
\label{recurrences}
\end{equation}
where~$\mathcal{U}_{\lambda+1}$ satisfies equation~(\ref{scalar mode eq})
with~$\lambda\!\to\! \lambda \!+\! 1$, and
where the proportionality constant was chosen for convenience. 
The Wronskian~(\ref{Wronskian or}) can consequently be expressed as,
\begin{equation}
{\rm Re} \Bigl[ \mathcal{U}_\lambda(\eta,k) \, \mathcal{U}_{\lambda+1}^*(\eta,k) \Bigr]
	= \frac{1}{2k} \, .
\label{Wronskian}
\end{equation}
Confirming the recurrence relations~(\ref{recurrences})
is accomplished by plugging them into the equation of motion~(\ref{scalar mode eq})
and applying condition~(\ref{time relation}) when commuting time derivatives.
In more special cases, such as power-law inflation~(\ref{power-law quants}), the
recurrence relations~(\ref{recurrences}) can be inferred from the properties of the 
solutions for the mode functions, such as recurrence
relations for Hankel functions that appear as solutions for the mode 
functions~(\ref{U power-law def}) (cf.~Sec.~2.2.1 from~\cite{Glavan:2022dwb}).
Recurrence relations~(\ref{recurrences}) help keep the expressions compact,
reducing the clutter of the computation.
They will be valid only for time-independent~$\zeta$ form the gauge-fixing term~(\ref{gauge-fixing term}),
but generalization to time-dependent values should be straightforward.

\subsection{Scalar two-point functions}
\label{subsec: Scalar two-point functions}

Nonequilibrium loop computations in quantum field theory require the use of several different 
two-point functions~\cite{Schwinger:1960qe,Mahanthappa:1962ex,Bakshi:1962dv,Bakshi:1963bn,Keldysh:1964ud,Chou:1984es,Calzetta:1986ey,Jordan:1986ug}.
The positive-frequency Wightman function can perhaps be considered the fundamental one,
as the remaining ones can all be expressed in terms of it. 
The Wightman function for scalar fields in FLRW 
can be expressed 
in terms of conformally rescaled scalar mode functions introduced in the preceding subsection
as a sum-over-modes,~\footnote{The sum-over-modes representation
needs to be regulated for it to be valid on the entire range of coordinates. 
This is accomplished by
appending an infinitesimal imaginary part to time 
coordinates: $\eta \!\to\! \eta \!-\! i\delta/2$, $\eta' \!\to\! \eta' \!-\! i\delta/2$.
Note that this substitution is first performed on the argument of the mode functions, and only then is
the complex conjugation in~(\ref{scalar two-point}) taken. In this sense, the two-point function
is defined as the distributional~$\delta\!\to\!0_+$ limit of an analytic function.}
\begin{equation}
i \bigl[ \tensor*[ ^{\scr \!-\!} ]{\Delta}{^{\scr \!+\! }} \bigr]_\lambda (x;x')
	= (aa')^{- \frac{D-2}{2}} \! \int\! \frac{ d^{D-1}k }{ (2\pi)^{D-1} } \,
	e^{i \vec{k} \cdot (\vec{x} - \vec{x}^{\,\prime} ) } \,
		\mathscr{U}_\lambda(\eta,\vec{k}) \mathscr{U}_{\lambda}^*(\eta',\vec{k})
		\, .
\label{scalar two-point}
\end{equation}
It satisfies the equation of motion inherited from the mode equation~(\ref{scalar mode eq}),
\begin{equation}
\Bigl[ \, \dalembertian
	- \bigl( \lambda_0^2 \!-\! \lambda^2 \bigr) (1\!-\!\epsilon)^2 H^2
	\Bigr] \,
	i \bigl[ \tensor*[^{\scr \!-\!}]{\Delta}{^{\scr \!+\!} } \bigr]_{\lambda} (x;x')
	= 0 \, ,
\label{scalar two-point function eom}
\end{equation}
where~$\dalembertian\!=\! g^{\mu\nu} \nabla_\mu \nabla_\nu$ is the 
d'Alembertian,~$\nabla_\mu$ is the~$D$-dimensional covariant derivative,
and
\begin{equation}
\lambda_0 = \frac{D \!-\! 1 \!-\! \epsilon}{ 2(1\!-\!\epsilon) } \, .
\end{equation}
The negative-frequency Wightman function is a complex conjugate of the positive-frequency 
one,~$i\bigl[ \tensor*[^{\scr \!+\!}]{\Delta}{^{\scr \!-\!}} \bigr]_\lambda(x;x') \!=\! 
	\bigl\{ i\bigl[ \tensor*[^{\scr \!-\!}]{\Delta}{^{\scr \!+\!}} \bigr]_\lambda(x;x') \bigr\}^*$.
The two Wightman functions can serve to define the Feynman propagator,
\begin{equation}
i \bigl[ \tensor*[^{\scr \!+\! }]{\Delta}{^{\scr \!+\!}} \bigr]_\lambda (x;x')
	=
	\theta(\eta\!-\!\eta') \, i \bigl[ \tensor*[^{\scr \!-\! }]{\Delta}{^{\scr \!+\!}} \bigr]_\lambda (x;x')
	+
	\theta(\eta'\!-\!\eta) \, i \bigl[ \tensor*[^{\scr \!+\! }]{\Delta}{^{\scr \!-\!}} \bigr]_\lambda (x;x')
	\, ,
\label{scalar Feynman def}
\end{equation}
which satisfies a sourced equation of motion,
\begin{equation}
\Bigl[ \dalembertian
	- \bigl( \lambda_0^2 \!-\! \lambda^2 \bigr) (1\!-\!\epsilon)^2 H^2
	\Bigr] \,
	i \bigl[ \tensor*[^{\scr \!+\!}]{\Delta}{^{\scr \!+\!} } \bigr]_{\lambda} (x;x')
	= \frac{ i \delta^{D}(x\!-\!x') }{ \sqrt{-g} } \, ,
\label{scalar Feynman eom}
\end{equation}
because of the step function~$\theta$ in its definition. Finally, the Dyson propagator is the complex conjugate 
of~(\ref{scalar Feynman def}),~$i\bigl[ \tensor*[^{\scr \!-\!}]{\Delta}{^{\scr \!-\!}} \bigr]_\lambda(x;x') \!=\! 
	\bigl\{ i\bigl[ \tensor*[^{\scr \!+\!}]{\Delta}{^{\scr \!+\!}} \bigr]_\lambda(x;x') \bigr\}^*$,
and it satisfies the equation of motion that is a conjugate of~(\ref{scalar Feynman eom}).
The sum-over-modes representations for different two-point functions
are inferred from the one for the Wightman function~(\ref{scalar two-point}) and their respective definitions.

\section{Classical photon in FLRW}
\label{sec: Classical photon in FLRW}

The massless vector field --- the photon ---  in general ~$D$-dimensional curved spacetimes is given by the
covariantized action for electromagnetism,
\begin{equation}
S[A_\mu] = \int\! d^{D\!}x \, \sqrt{-g} \, \biggl[
	- \frac{1}{4} g^{\mu\rho} g^{\nu\sigma} F_{\mu\nu} F_{\rho\sigma}
	\biggr] \, ,
\label{invariant action}
\end{equation}
where~$F_{\mu\nu}\!=\! \partial_\mu A_\nu \!-\! \partial_\nu A_\mu$ is the field strength tensor
for the vector potential~$A_\mu$. This action is invariant under~$U(1)$ gauge 
transformations,~$A_\mu \!\to\! A_\mu \!+\! \partial_\mu \Lambda$, where~$\Lambda(x)$ is
an arbitrary scalar function. Consequently the covariant Maxwell's equations 
inherit this property, meaning their solutions are not
fully determined by specifying initial conditions for the vector potential components. 
This implies that the number of physical propagating degrees of freedom is smaller than the number of 
dynamical fields. Just as in flat space, the propagating degrees of freedom of the 
free photon 
are the~$(D\!-\!2)$ transverse polarizations. Their dynamics is well understood, since 
in~$D\!=\!4$ they couple conformally to gravity
and effectively do not see the expansion. However, formulating interacting gauge theories in terms of physical
propagating degrees of freedom only is rather impractical at best. It is 
often advantageous to consider the questions of
the dynamics and of observables separately.
One first fixes the gauge to remove the ambiguities in the dynamical equations, 
and solves for the gauge-fixed dynamics. Then one obtains observables by projecting out the
physical information from the gauge-fixed solutions.

Particularly convenient gauges for loop computations in quantum field theory are the so-called
{\it average gauges}, known as {\it multiplier gauges} in classical theory~\cite{Henneaux:1992ig}.
These are not imposed by
following the Dirac-Bergmann algorithm~\cite{Dirac:2011} and imposing gauge conditions that eliminate 
some of the vector potential components. Instead, multiplier gauges treat all the components of the
vector potential on equal footing, and fix the ambiguities in the dynamics by fixing directly the
Lagrange multipliers. This procedure is ultimately equivalent to 
adding a gauge-fixing term to the gauge-invariant action~(\ref{invariant action}).  A specific choice for 
the multiplier will lead to the gauge-fixing term in~(\ref{gauge-fixing term}) we consider here.
This sort of gauge fixing is not often encountered in classical theories, perhaps giving the
impression that there is something innately quantum about gauge-fixing terms added to gauge-invariant actions. 
This is far from true, as the rationale behind gauge-fixing terms is essentially
the classical canonical structure in multiplier gauges. 
That is why this section is devoted to presenting the details of implementing multiplier gauges,
in particular for the photon in~(\ref{invariant action}). Canonical formulation in multiplier gauges
subsequently lends itself readily to canonical quantization that is considered in Sec.~\ref{sec: Quantized photon in FLRW}.

\subsection{Gauge-invariant canonical formulation}
\label{subsec: Gauge-invariant canonical formulation}

Our starting point of implementing a multiplier gauge
is the canonical formulation of the action~(\ref{invariant action})
that we derive in this subsection. We start by decomposing the indices in~(\ref{invariant action}) into
spatial and temporal ones,
\begin{equation}
S[A_\mu] = \int\! d^{D\!}x \, a^{D-4} \, \biggl[
	\frac{1}{2} F_{0i} F_{0i}
	- \frac{1}{4} F_{ij} F_{ij}
	\biggr]
	\, ,
\label{decomposed indices}
\end{equation}
where henceforth all such decomposed indices are written as lowered, 
and the convention that
repeated spatial indices are summed over is implied.
We follow~\cite{Gitman} in
reformulating~(\ref{decomposed indices}) as a first order
canonical action. This requires first
promoting all the time derivatives to independent velocity
fields,
\begin{equation}
\partial_0 A_0 \to V_0 \, ,
\qquad \qquad
F_{0i} = \partial_0 A_i - \partial_i A_0 \to V_i \, ,
\end{equation}
and introducing accompanying Lagrange multipliers~$\Pi_0$ and~$\Pi_i$ that ensure
on-shell equivalence of the intermediate extended action,
\begin{align}
\MoveEqLeft[4]
\mathcal{S}\bigl[ A_0, V_0, \Pi_0, A_i, V_i, \Pi_i \bigr] 
	= \int\! d^{D\!}x \, 
	\biggl\{ a^{D-4} \biggl[
	\frac{1}{2} V_i V_i
	- \frac{1}{4} F_{ij} F_{ij}
	\biggr]
\nonumber \\
&	\hspace{3.5cm}
	+ \Pi_0 \Bigl( \partial_0 A_0 - V_0  \Bigr)
	+ \Pi_i \Bigl( \partial_0 A_i - \partial_i A_0 - V_i \Bigr)
	\biggr\}  \, ,
\end{align}
to the original action~(\ref{invariant action}).
Solving for the velocity fields on-shell, which here is possible only for~$V_i$,
\begin{equation}
\frac{\delta \mathcal{S}}{ \delta V_i}
	\approx 0 \, ,
\qquad \Longrightarrow \qquad
V_i \approx \overline{V}_i = a^{4-D} \Pi_i \, .
\end{equation}
and plugging this into the extended action above produces the canonical action,
\begin{align}
\mathscr{S}\bigl[ A_0, \Pi_0, A_i, \Pi_i, \ell \, \bigr]
	\equiv{}& \mathcal{S}\bigl[ A_0, V_0 \!\to\! \ell , \Pi_0, A_i, \overline{V}_i, \Pi_i \bigr] 
\nonumber \\
&
	=
	\int\! d^{D\!}x \, \Bigl[ \Pi_0 \partial_0 A_0 + \Pi_i \partial_0 A_i - \mathscr{H} - \ell \, \Psi_1 \Bigr] \, ,
\label{canonical action}
\end{align}
where
\begin{equation}
\mathscr{H} = \frac{a^{4-D}}{2} \Pi_i \Pi_i
	+ \Pi_i \partial_i A_0
	+ \frac{a^{D-4}}{4} F_{ij} F_{ij}
\end{equation}
is the canonical Hamiltonian density and~$\Psi_1\!=\!\Pi_0$ is the primary constraint generated by the
Lagrange multiplier~$\ell$ (which is just a relabeled field~$V_0$).
Variation of~(\ref{canonical action}) with respect to the canonical variables produces 
canonical equations of motion,
\begin{equation}
\partial_0 A_0 \approx  \ell \, ,
\qquad
\partial_0 \Pi_0 \approx
	\partial_i \Pi_i  \, ,
\qquad
\partial_0 A_i \approx
	a^{4-D} \Pi_i + \partial_i A_0
	\, ,
\qquad
\partial_0 \Pi_i \approx a^{D-4} \partial_j F_{ji} \, ,
\label{invariant EOMs}
\end{equation}
while variation with respect to the Lagrange multiplier~$\ell$ gives the primary constraint,
\begin{equation}
\Psi_1 \approx 0 \, .
\label{primary constraint}
\end{equation}
Note that in this section we employ Dirac's notation where~$=$ stands for an off-shell (strong) equality,
while~$\approx$ stands for an on-shell (weak) equality.
The equations above are equivalent to Hamilton's equations 
descending from the total Hamiltonian~$\mathscr{H}_{\rm tot} \!=\! \mathscr{H}\!+\! \ell \, \Psi_1$,
where the Poisson brackets are determined
by the symplectic part of the canonical action~(\ref{canonical action}), with 
the nonvanishing ones being
\begin{equation}
\bigl\{ A_0(\eta,\vec{x}) , \Pi_0(\eta,\vec{x}^{\,\prime}) \bigr\} = \delta^{D-1}(\vec{x} \!-\! \vec{x}^{\,\prime}) \, ,
\qquad
\bigl\{ A_i(\eta,\vec{x}) , \Pi_j(\eta,\vec{x}^{\,\prime}) \bigr\} = \delta_{ij} \delta^{D-1}(\vec{x} \!-\! \vec{x}^{\,\prime}) \, ,
\label{brackets}
\end{equation}
and the Lagrange multiplier~$\ell$ has vanishing brackets with all the canonical variables.
Note that the constraint equation~(\ref{primary constraint}) does not follow from the total Hamiltonian, 
but needs to be considered in addition to Hamilton's equations.

The consistency of the primary constraint~(\ref{primary constraint}) requires it to be conserved,
which in turn generates a secondary constraint,
\begin{equation}
0 \approx \partial_0 \Psi_1 \approx
	\partial_i \Pi_i  \equiv \Psi_2 \, ,
\end{equation}
the conservation of which generates no further constraints,
\begin{equation}
\partial_0 \Psi_2 \approx 0 \, .
\label{secondary constraint}
\end{equation}
The primary and secondary constraints form a complete set of first-class constraints,
\begin{equation}
\bigl\{ \Psi_1(\eta,\vec{x}) , \Psi_2(\eta,\vec{x}^{\,\prime}) \bigr\} = 0 \, ,
\end{equation}
implying that the Lagrange multiplier~$\ell$ is left undetermined by the equations
of motion, which is how gauge symmetries manifest themselves in the canonical formulation.
Fixing this ambiguity is what the next section is devoted to.

While any solution to the dynamical equations~(\ref{invariant EOMs})
and the accompanying constraint equations~(\ref{primary constraint}) and~(\ref{secondary constraint})
describes the same physical system, observables in gauge theories cannot
depend on the arbitrary Lagrange multipliers such as~$\ell$. This is guaranteed by 
requiring observables to have on-shell vanishing
brackets with all the first-class constraints. For the case at hand, this means that
for some~$\mathscr{O}(\eta,\vec{x})$ to be an observable it has to satisfy
\begin{equation}
\bigl\{ \Psi_1(\eta,\vec{x}) , \mathscr{O}(\eta,\vec{x}^{\,\prime}) \bigr\}
	\approx 0 \, ,
	\qquad \qquad
\bigl\{ \Psi_2(\eta,\vec{x}) , \mathscr{O}(\eta,\vec{x}^{\,\prime}) \bigr\}
	\approx 0 \, ,
\label{class observable}
\end{equation}
which guarantees that it does not depend on the arbitrary Lagrange multiplier~$\ell$.

\subsection{Gauge-fixed canonical formulation}
\label{subsec: Gauge-fixed canonical formulation}


There are multiple ways of fixing the ambiguity of the dynamical equations~(\ref{invariant EOMs}) 
of the preceding section. The Dirac-Bergman algorithm requires the specification of gauge conditions
in the form of off-shell equalities that eliminate part of the dynamical canonical fields. Such gauge choices
are known as {\it exact gauges}, and the Coulomb gauge is one classic example. However, exact gauges are often 
impractical to use in quantized theories. Preferred choices
are the so-called multiplier gauges (also called average gauges) that do not eliminate
any of the dynamical fields, but rather treat all of them on an equal footing.
Implementing a multiplier gauge for the problem at hand
is accomplished by fixing by hand the multiplier~$\ell$ to be a function of canonical pairs,
without explicitly specifying any gauge conditions,
\begin{equation}
\ell \to \overline{\ell} \bigl( A_0, \Pi_0, A_i, \Pi_i \bigr) \, .
\label{L choice}
\end{equation}
%
%
%
Employing this choice at the level of equations of motion~(\ref{invariant EOMs})  
produces a set of gauge-fixed equations of motion,
\begin{align}
\partial_0 A_0 \approx{}&
	\overline{\ell} \bigl( A_0, \Pi_0, A_i, \Pi_i \bigr) \, ,
\label{on-shell eq1}
\\
\partial_0 \Pi_0 \approx{}& 
	\partial_i \Pi_i 
\label{on-shell eq2}
\\
\partial_0 A_i \approx{}& 
	a^{4-D} \Pi_i + \partial_i A_0 \, ,
\label{on-shell eq3}
\\
\partial_0 \Pi_i \approx{}& 
	a^{D-4} \partial_j F_{ji} \, ,
\label{on-shell eq4}
\end{align}
%
%
%
in addition to the two first-class constraints that remain unchanged,
\begin{equation}
\Psi_1 = \Pi_0 \approx 0 \, ,
\qquad \qquad
\Psi_2 = \partial_i \Pi_i \approx 0 \, .
\label{on-shell eq5}
\end{equation}
This system of equations now forms a well-defined initial value problem (provided that~$\overline{\ell}$
is not chosen in some pathological manner). 
A useful way of viewing these equations is to consider the two first-class constraints as 
conditions on the initial value surface. The evolution will ensure they are preserved for all times.
This way we split the problem into four dynamical equations~(\ref{on-shell eq1})--(\ref{on-shell eq4}) 
describing evolution,
and two kinematic equations~(\ref{on-shell eq5}) constraining the initial conditions. The latter
cut define a subspace of the space of solutions of the the former.
This shows that choosing~$\ell \!\to\! \overline{\ell}$ at the level of the equations of motion
leads to a well defined dynamical problem and in that sense it fixes the gauge.
However, it is more convenient to implement this gauge at the level of the action.

Instead of fixing the multiplier at the level of equations of motion, we can 
fix it in the canonical action~(\ref{canonical action}) directly. This defines the {\it gauge-fixed canonical
action},
\begin{equation}
\mathscr{S}_{\star}\bigl[ A_0, \Pi_0, A_i, \Pi_i \bigr]
	\equiv \mathscr{S}\bigl[ A_0, \Pi_0, A_i, \Pi_i, \ell\!\to\! \overline{\ell} \, \bigr]
	=
	\int\! d^{D\!}x \, \Bigl[
		\Pi_0 \partial_0 A_0 + \Pi_i \partial_0 A_i - \mathscr{H}_\star
		\Bigr]
		\, ,
\label{gauge-fixed canonical action}
\end{equation}
where the gauge-fixed Hamiltonian density is
\begin{equation}
\mathscr{H}_{\star} = \mathscr{H} + \overline{\ell} \, \Psi_1 \, .
\end{equation}
Note that this gauge-fixed action is not equivalent to canonical action~(\ref{canonical action}) as it no
longer encodes a variational principle with respect to the multiplier~$\ell$.
The equations of motion generated by the gauge-fixed canonical action 
are~\footnote{Partial derivatives of~$\overline{\ell}$ are generalized to
functional derivatives in an obvious way whenever necessary. This detail is omitted for notational simplicity.}
\begin{align}
\partial_0 A_0 \approx{}&
	\overline{\ell} 
	+ \frac{\partial \overline{\ell} }{ \partial \Pi_0 } \Pi_0
	\, ,
\label{gf A0 eq}
\\
\partial_0 \Pi_0 \approx{}& 
	\partial_i \Pi_i 
	- \frac{\partial \overline{\ell} }{ \partial A_0 } \Pi_0
	\, ,
\label{gf Pi0 eq}
\\
\partial_0 A_i \approx{}& 
	a^{4-D} \Pi_i + \partial_i A_0 
	+ \frac{\partial \overline{\ell} }{ \partial \Pi_i } \Pi_0
	\, ,
\label{gf Pi0 eq}
\\
\partial_0 \Pi_i \approx{}& 
	a^{D-4} \partial_j F_{ji} 
	- \frac{\partial \overline{\ell} }{ \partial A_i } \Pi_0 
	\, .
\label{gf Pii eq}
\end{align}
This is now a fully determined set of coupled partial differential equations in the sense
that specifying initial conditions fully fixes the evolution.
However, note that these are not immediately equivalent to 
Eqs.~(\ref{on-shell eq1})--(\ref{on-shell eq5}). 
First, the four dynamical equations are all modified by additional
terms, and more importantly the constraint equations are absent. 
The remedy is to consider the gauge-fixed action to encode the dynamics only, 
and to require the first-class constraints (37) to be satisfied {\it in addition} to the 
dynamical equations by considering them to be {\it subsidiary conditions} on the initial 
value surface,
\begin{equation}
\Psi_1(\eta_0,\vec{x}) = \Pi_0(\eta_0,\vec{x}) \approx 0 \, ,
\qquad \qquad
\Psi_2(\eta_0,\vec{x}) = \partial_i \Pi_i(\eta_0,\vec{x}) \approx 0 \, .
\label{const eq}
\end{equation}
Equations of motion~(\ref{gf A0 eq})--(\ref{gf Pii eq}) guarantee that imposing constraints~(\ref{const eq})
on the initial value surface is sufficient to guarantee their conservation.
%
%
The system of Eqs.~(\ref{gf A0 eq})--(\ref{const eq}) is now obviously equivalent to the original 
system~(\ref{on-shell eq1})--(\ref{on-shell eq5}).

\medskip

The choice for the multiplier in~(\ref{L choice}) is not dictated by physical principles,
but is rather a matter of convenience, as in fact any gauge choice is. In this work we consider
only linear gauges natural for free theories. We require them (i) to respect homogeneity 
and isotropy of the FLRW spacetime, (ii) not to introduce additional dimensionful scales,
(iii) to be composed of commensurate terms, 
and (iv) to respect Lorentz invariance in the Minkowski limit.
This essentially restricts the choice 
of the multiplier to a two-parameter family,
\begin{equation}
\overline{\ell} = - \frac{\xi}{2} a^{4-D} \Pi_0 
	+ \partial_i A_i - (D\!-\!2\!-\!2\zeta) \mathcal{H} A_0
	\, ,
\label{ell bar}
\end{equation}
where~$\xi$ and~$\zeta$ are two arbitrary dimensionless gauge-fixing 
parameters.\footnote{Having~$\xi$ and~$\zeta$ 
be time-dependent functions would be just as easy, but we do not consider it for simplicity;
the generalization is straightforward.}
This particular gauge choice then produces the equations of motion,
\begin{align}
\partial_0 A_0 \approx{}&
	- \xi a^{4-D} \Pi_0
	+ \partial_i A_i 
	- (D\!-\!2\!-\!2\zeta) \mathcal{H} A_0
	\, ,
\label{fixed EOM1}
\\
\partial_0 \Pi_0 \approx{}& 
	\partial_i \Pi_i 
	+ (D \!-\! 2 \!-\! 2\zeta) \mathcal{H} \Pi_0
	\, ,
\label{fixed EOM2}
\\
\partial_0 A_i \approx{}& 
	a^{4-D} \Pi_i + \partial_i A_0 
	\, ,
\label{fixed EOM3}
\\
\partial_0 \Pi_i \approx{}& 
	\partial_i \Pi_0
	+ a^{D-4} \partial_j F_{ji} 
	\, ,
\label{fixed EOM4}
\end{align}
that are generated by the gauge-fixed Hamiltonian
\begin{equation}
\mathscr{H}_\star = 
	\frac{a^{4-D}}{2} \bigl( \Pi_i \Pi_i
	- \xi \Pi_0 \Pi_0 \bigr)
	+ \Pi_i \partial_i A_0
	+ \Pi_0 \partial_i A_i
	- (D\!-\!2\!-\!2\zeta) \mathcal{H} \Pi_0  A_0
	+ \frac{a^{D-4}}{4} F_{ij} F_{ij}
	\, .
\label{gf Hamiltonian}
\end{equation}
The two first-class constraints~(\ref{const eq}) satisfy closed equations of motion,
\begin{equation}
\partial_0 \Psi_1 \approx \Psi_2 + (D\!-\!2\!-\!2\zeta) \mathcal{H} \Psi_1 \, ,
\qquad \qquad
\partial_0 \Psi_2 \approx \nabla^2 \Psi_1 \, ,
\end{equation}
where~$\nabla^2\!\equiv\!\partial_i\partial_i$ is the Laplace operator,
exemplifying the fact that the dynamics preserves~(\ref{const eq}) 
if imposed on the initial value surface.

We have defined the gauge-fixed system in terms of the gauge-fixed canonical 
action~(\ref{gauge-fixed canonical action}) with the choice for the multiplier in~(\ref{ell bar}) describing the
dynamics, and the two subsidiary conditions~(\ref{const eq}) accounting for first-class constraints.
Despite being the superior formulation for analyzing the structure of gauge theories, the
canonical formulation is often less intuitive than the configurations space formulation.
There is an associated gauge-fixed configuration space action associated 
with the canonical one.
By solving for the canonical momenta on-shell,
\begin{align}
\frac{\delta \mathscr{S}_\star }{ \delta \Pi_0}
	\approx{}& 0
\ \  \Longrightarrow \ \ 
	\Pi_0 \approx \overline{\Pi}_0 
		= - \frac{a^{D-4}}{ \xi } 
			\Bigl( \partial_0 A_0 + (D\!-\!2 \!-\! 2\zeta) \mathcal{H} A_0 
				- \partial_i A_i
				 \Bigr)
			\, ,
\\
\frac{\delta \mathscr{S}_\star }{ \delta \Pi_i}
	\approx{}& 0
\ \  \Longrightarrow \ \ 
	\Pi_i \approx \overline{\Pi}_i
		= a^{D-4} \Bigl( \partial_0 A_i - \partial_i A_0 \Bigr) 
		\, .
\end{align}
and inserting the solutions into the gauge-fixed canonical action produces 
precisely the associated gauge-fixed configuration space action
\begin{equation}
S_\star[A_\mu] = \int\! d^{D\!}x \, \sqrt{-g} \, \biggl[
	- \frac{1}{4} g^{\mu\rho} g^{\nu\sigma} F_{\mu\nu} F_{\rho\sigma}
	- \frac{1}{2\xi} \Bigl( \nabla^\mu A_\mu - 2 \zeta n^\mu A_\mu \Bigr)^{\!2\,}
	\biggr] \, .
\label{fixed action}
\end{equation}
with the gauge-fixing term~(\ref{gauge-fixing term}).
Thus we have derived how the gauge-fixing terms arise in classical gauge theories.
The subsidiary conditions~(\ref{const eq}) in the configuration space formulation 
take the form
\begin{equation}
\xi a^{2-D} \Psi_1 \approx
\bigl( \nabla^\mu - 2 \zeta n^\mu \bigr) A_\mu \approx 0 \, ,
\qquad \qquad
a^{4-D}\Psi_2
\approx \partial_i F_{0i} \approx 0 \, .
\label{configuration space constraints}
\end{equation}
%


%
%

\section{Quantized photon in FLRW}
\label{sec: Quantized photon in FLRW}

The classical theory in the multiplier gauge is formulated so that the dynamics is represented by 
a gauge-fixed action without constraints, and the first-class constraints are imposed 
as subsidiary conditions on
the initial value surface. The canonical quantization of such gauge-fixed theory is most naturally 
implemented in
the Heisenberg picture, where field operators account for the dynamics, and the state vector
accounts for the initial conditions.

\subsection{Dynamics}
\label{sec: Dynamics}

The usual rules of canonical quantization are sufficient to quantize the gauge-fixed dynamics
of the classical theory. Canonical fields are promoted to Hermitian field operators,
\begin{equation}
A_0(x) \to \hat{A}_0(x) \, ,
\qquad
\Pi_0(x) \to \hat{\Pi}_0(x) \, ,
\qquad
A_i(x) \to \hat{A}_i(x) \, ,
\qquad
\Pi_i(x) \to \hat{\Pi}_i(x) \, ,
\end{equation}
and their Poisson brackets~(\ref{brackets}) to commutators,
\begin{equation}
\bigl[ \hat{A}_0(\eta,\vec{x}) , \hat{\Pi}_0(\eta,\vec{x}^{\,\prime}) \bigr] 
	= i \delta^{D-1}(\vec{x} \!-\! \vec{x}^{\,\prime}) \, ,
\qquad
\bigl[ \hat{A}_i(\eta,\vec{x}) , \hat{\Pi}_j(\eta,\vec{x}^{\,\prime}) \bigr] 
	= \delta_{ij} \, i \delta^{D-1}(\vec{x} \!-\! \vec{x}^{\,\prime}) \, .
\label{ccr}
\end{equation}
The gauge-fixed equations of motion~(\ref{fixed EOM1})--(\ref{fixed EOM4}) remain unchanged,,
\begin{align}
\partial_0 \hat{A}_0
	={}&
	- \xi a^{4-D} \hat{\Pi}_0 + \partial_i \hat{A}_i - (D\!-\!2\!-\!2\zeta) \mathcal{H} \hat{A}_0
	\, ,
\label{operator eom 1}
\\
\partial_0 \hat{\Pi}_0
	={}&
	\partial_i \hat{\Pi}_i + (D\!-\!2 \!-\! 2\zeta) \mathcal{H} \hat{\Pi}_0
	\, ,
\label{operator eom 2}
\\
\partial_0 \hat{A}_i
	={}&
	a^{4-D} \hat{\Pi}_i + \partial_i \hat{A}_0
	\, ,
\label{operator eom 3}
\\
\partial_0 \hat{\Pi}_i
	={}&
	\partial_i \hat{\Pi}_0 + a^{D-4} \partial_j \hat{F}_{ji}
	\, ,
\label{operator eom 4}
\end{align}
and are generated by the Hamiltonian~(\ref{gf Hamiltonian}) with fields promoted to field operators.

\subsection{Subsidiary condition}
\label{sec: Subsidiary condition}

Implementing the first-class constraints in the quantized theory is less straightforward.
It definitely must involve Hermitian constraint operators,
\begin{equation}
\hat{\Psi}_1(x) = \hat{\Pi}_0(x) \, ,
\qquad \qquad
\hat{\Psi}_2(x) = \partial_i \hat{\Pi}_i (x) \, .
\label{Hermitian const}
\end{equation}
However, the constraints cannot be imposed as operator equalities as that would contradict
canonical commutation relations~(\ref{ccr}). 
To understand how to quantize the first-class constraints we better first consider 
the correspondence principle, which tells us that matrix elements of Hermitian first-class 
constraints have to vanish at initial time,
\begin{equation}
\bigl\langle \Omega_1 \bigr| \hat{\Psi}_1(\eta_0,\vec{x}) \bigr| \Omega_2 \bigr\rangle = 0 \, ,
\qquad \qquad
\bigl\langle \Omega_1 \bigr| \hat{\Psi}_2(\eta_0,\vec{x}) \bigr| \Omega_2 \bigr\rangle = 0 \, .
\label{expectation condition}
\end{equation}
This cannot be satisfied by requiring that the Hermitian constraints themselves annihilate the 
state vector (as required in~\cite{Vollick:2012cn,Cresswell:2015vaa}), as that would again contradict 
the canonical commutation relations~(\ref{ccr}).
However, it is consistent to require that the state ket-vector is annihilated by an invertible
non-Hermitian linear combination of the two constraints~(\ref{Hermitian const}),
\begin{equation}
\hat{K}(\vec{x}) = \int\! d^{D-1} x' \,
	\Bigl[ f_1(\eta_0, \vec{x}\!-\!\vec{x}^{\,\prime}) \hat{\Psi}_1(\eta_0,\vec{x}^{\,\prime} ) 
		+ f_2(\eta_0, \vec{x}\!-\!\vec{x}^{\,\prime}) \hat{\Psi}_2(\eta_0,\vec{x}^{\,\prime} ) 
		\Bigr] \, ,
\label{Kdef}
\end{equation}
that we refer to as a {\it subsidiary non-Hermitian constraint operator}, 
and that its conjugate annihilates the state bra-vector,
\begin{equation}
\hat{K}(\vec{x}) \bigl| \Omega \bigr\rangle = 0 \, ,
\qquad \quad
\bigl\langle \Omega \bigr| \hat{K}^\dag(\vec{x}) = 0 \, ,
\qquad \quad
\forall \vec{x} \, .
\label{K position condition}
\end{equation}
This condition is a generalization of the Gupta-Bleuler subsidiary condition for covariant 
photon gauges in flat space. It is divorced from spacetime symmetries and symmetries of the 
gauge-fixing term, and is rather based solely on the canonical structure.

The conditions in~(\ref{expectation condition}) are preserved in time,
\begin{equation}
\bigl\langle \Omega_1 \bigr| \hat{\Psi}_1(\eta,\vec{x}) \bigr| \Omega_2 \bigr\rangle = 0 \, ,
\qquad \qquad
\bigl\langle \Omega_1 \bigr| \hat{\Psi}_2(\eta,\vec{x}) \bigr| \Omega_2 \bigr\rangle = 0 \, ,
\end{equation}
due to the equations of motion the Hermitian constraints satisfy,
\begin{equation}
\partial_0 \hat{\Psi}_1
	=
	\hat{\Psi}_2 + (D\!-\!2 \!-\! 2\zeta) \mathcal{H} \hat{\Psi}_1
	\, ,
\qquad \qquad
\partial_0 \hat{\Psi}_2
	=
	\nabla^2 \hat{\Psi}_1
	\, .
\label{Her motion}
\end{equation}
This implies that the time-independent non-Hermitian constraint can be expressed in
terms of Hermitian ones at any point in time,
since there are~$f_1$ and~$f_2$ at any time~$\eta$ such that,
\begin{equation}
\hat{K}(\vec{x}) = \int\! d^{D-1} x' \,
	\Bigl[ f_1(\eta, \vec{x}\!-\!\vec{x}^{\,\prime}) \hat{\Psi}_1(\eta,\vec{x}^{\,\prime} ) 
		+ f_2(\eta, \vec{x}\!-\!\vec{x}^{\,\prime}) \hat{\Psi}_2(\eta,\vec{x}^{\,\prime} ) 
		\Bigr] \, .
\label{K position definition}
\end{equation}
Given the equations of motion~(\ref{Her motion}),
the coefficient functions have to satisfy,
\begin{equation}
\partial_0 f_1 = - \nabla^2 f_2 - (D\!-\!2\!-\!2\zeta) \mathcal{H} f_1 \, ,
\qquad \qquad
\partial_0 f_2 = - f_1 \, .
\end{equation}

The subsidiary constraint operator~(\ref{Kdef}) commutes with its conjugate,
\begin{equation}
\bigl[ \hat{K}(\vec{x}) , \hat{K}^\dag(\vec{x}^{\,\prime}) \bigr] = 0 \, ,
\end{equation}
and consequently a matrix element of any polynomial functional of Hermitian constraints 
vanishes,
\begin{equation}
\bigl\langle \Omega_1 \bigr| \mathscr{P}
	\bigl[ \hat{\Psi}_1(\eta,\vec{x}) , \hat{\Psi}_2(\eta,\vec{x}) \bigr]
	\bigl| \Omega_2 \bigr\rangle
	= 0 \, ,
\end{equation}
as required by the correspondence principle. In particular, the two-point functions of 
Hermitian constraints must vanish,
\begin{subequations}
\begin{align}
\bigl\langle \Omega \bigr| \hat{\Psi}_1(\eta,\vec{x}) \hat{\Psi}_1(\eta',\vec{x}^{\,\prime}) 
	\bigl| \Omega \bigr\rangle ={}& 0 \, , 
\\
\bigl\langle \Omega \bigr| \hat{\Psi}_1(\eta,\vec{x}) \hat{\Psi}_2(\eta',\vec{x}^{\,\prime}) 
	\bigl| \Omega \bigr\rangle ={}& 0 \, , 
\\
\bigl\langle \Omega \bigr| \hat{\Psi}_2(\eta,\vec{x}) \hat{\Psi}_2(\eta',\vec{x}^{\,\prime}) 
	\bigl| \Omega \bigr\rangle ={}& 0 \, .
\end{align}
\label{Hermitian two-points}%
\end{subequations}

For later sections it is useful to define some shorthand notation. Namely, we can invert 
relation~(\ref{K position definition}) 
and express the Hermitian constraints in terms of the non-Hermitian ones,
\begin{equation}
\hat{\Psi}_1(x) = \hat{K}_1^\dag(x) + \hat{K}_1(x) \, ,
\qquad \qquad
\hat{\Psi}_2(x) = \hat{K}_2^\dag(x) + \hat{K}_2(x) \, ,
\label{Hermitian decomposition}
\end{equation}
where pieces~$\hat{K}_1$ and~$\hat{K}_2$ contain just~$\hat{K}$, and their conjugates contain
just~$\hat{K}^\dag$, such that for any physical state we have
\begin{equation}
\hat{K}_1(x) \bigl| \Omega \bigr\rangle
	= \hat{K}_2(x) \bigl| \Omega \bigr\rangle
	= 0
	\, ,
\qquad \qquad
\bigl\langle \Omega \bigr| \hat{K}_1^\dag(x)
	= \bigl\langle \Omega \bigr| \hat{K}_2^\dag(x)
	= 0
	\, .
\label{K12 properties}
\end{equation}
%

\subsection{Quantum observables}
\label{subsec: Quantum observables}

A classical observable~$\mathscr{O}$ is a quantity composed out of canonical fields
that has vanishing Poisson brackets with all the first-class constraints~(\ref{class observable}).
When promoting such a classical observable to a quantum observable we have
to promote~$\mathscr{O}$ to an operator~$\hat{\mathscr{O}}$, 
by promoting the canonical fields it is composed of to canonical field operators. 
This process requires that we address the question of operator ordering 
in~$\hat{\mathscr{O}}$.

\medskip

In quantum gauge theories there is an additional operator ordering issue compared to 
quantum theories without constraints.
To understand this ordering issue consider a trivial classical 
``observable,''~\footnote{We consider for simplicity a product of two fields evaluated 
at different spatial points, to avoid having to multiply distributions in the quantized theory.}
\begin{equation}
\mathscr{O}(\eta,\vec{x},\vec{x}^{\,\prime})
	=
	A_0(\eta,\vec{x}) \Psi_2(\eta,\vec{x}^{\,\prime})
	\approx 0 \, ,
\label{trivial observable}
\end{equation}
that vanishes on account of being proportional to one of the first-class 
constraints~(\ref{class observable}).
When promoting this quantity to an operator we may first consider Weyl ordering
(denoted by subscript~$\rm W$ henceforth). 
In this particular case there is no
need for explicit Weyl ordering as the involved field operators commute,
\begin{equation}
\bigl[ \hat{\mathscr{O}}(\eta,\vec{x}, \vec{x}^{\,\prime}) \bigr]_{\scr \rm W}
	\equiv
	\frac{1}{2} \Bigl( \hat{A}_0(\eta,\vec{x}) \hat{\Psi}_2(\eta,\vec{x}^{\,\prime}) 
	+ \hat{\Psi}_2(\eta,\vec{x}^{\,\prime}) \hat{A}_0(\eta,\vec{x}) \Bigr)
	=
	\hat{A}_0(\eta,\vec{x}) \hat{\Psi}_2(\eta,\vec{x}^{\,\prime})
	\, .
\end{equation}
However, taking the expectation value produces a nonvanishing result,
\begin{equation}
\bigl\langle \Omega \bigr| \bigl[ \hat{\mathscr{O}}(\eta,\vec{x}, \vec{x}^{\,\prime}) \bigr]_{\scr \rm W} \bigl| \Omega \bigr\rangle
	=
	{\rm Re} \Bigl( \big[ \hat{K}_2(\eta,\vec{x}^{\,\prime}) , \hat{A}_0(\eta,\vec{x}) \bigr] \Bigr)
	\, ,
\label{nonv}
\end{equation}
where we use the shorthand notation of the decomposition in~(\ref{Hermitian decomposition}). This is independent of  any physical state 
satisfying~(\ref{K position condition}), as it is only the commutator 
that appears on the right-hand side of~(\ref{nonv}).
This nonvanishing expectation value violates the correspondence principle;
even though the right-hand side does not depend on the quantum state, the choice
of~$\hat{K}$ is still largely arbitrary, and consequently so is~$\hat{K}_2$. The proper way to order the operators 
(henceforth denoted by subscript~$\rm g$ for ``gauge'')
in observable~(\ref{trivial observable}) is to 
(i) decompose the Hermitian constraint operator~$\hat{\Psi}_2$
into the non-Hermitian subsidiary constraint operator~$\hat{K}$ and its
conjugate~$\hat{K}^\dag$, and (ii) put all~$\hat{K}$ operators
on the right of the product, and all~$\hat{K}^\dag$ operators on 
the left of the product,
\begin{equation}
\bigl[ \hat{\mathscr{O}}(\eta,\vec{x}, \vec{x}^{\,\prime}) \bigr]_{\rm g}
	=
	\hat{K}_2^\dag(\eta,\vec{x}^{\,\prime}) \hat{A}_0(\eta,\vec{x})
	+
	\hat{A}_0(\eta,\vec{x}) \hat{K}_2(\eta,\vec{x}^{\,\prime})
	\, .
\label{gauge ordered O}
\end{equation}
Such ordering guarantees that the expectation value vanishes due to~(\ref{K12 properties}) for any 
physical state,
\begin{equation}
\bigl\langle \Omega \bigr| \bigl[ \hat{\mathscr{O}}(\eta,\vec{x}, \vec{x}^{\,\prime}) 
	\bigr]_{\rm g} \bigl| \Omega \bigr\rangle
	=
	0
	\, .
\label{g ordering}
\end{equation}
It is useful to note that the properly ordered observable can be written in terms of
the Weyl-ordered observable plus the commutator accounting for the difference,
\begin{equation}
\bigl[ \hat{\mathscr{O}}(\eta,\vec{x}, \vec{x}^{\,\prime}) \bigr]_{\rm g}
	=
	\bigl[ \hat{\mathscr{O}}(\eta,\vec{x}, \vec{x}^{\,\prime}) \bigr]_{\scr \rm W}
	-
	{\rm Re} \Bigl( \big[ \hat{K}_2(\eta,\vec{x}^{\,\prime}) , \hat{A}_0(\eta,\vec{x}) \bigr] \Bigr)
	\, .
\label{Og to W}
\end{equation}
We comment more on the significance of this in Sec.~\ref{subsec: Energy-momentum tensor}.
The difference between two orderings is ultimately the Faddeev-Popov ghost contribution.

Having addressed the operator ordering in~(\ref{gauge ordered O}), we need to consider further quantum
properties of this trivial observable, which are usually encoded by correlators ($n$-point functions).
Since it contains a Hermitian
constraint and vanishes trivially classically,  its~$n$-point functions ought to vanish as well.
However, while the operator ordering in~(\ref{g ordering}) guarantees this is satisfied for
the expectation values (one-point function),
the two-point function of this operator does not vanish,
\begin{equation}
\bigl\langle \Omega \bigr| \bigl[ \hat{\mathscr{O}}(\eta,\vec{x}, \vec{x}^{\,\prime}) 
	\bigr]_{\rm g}
	\bigl[ \hat{\mathscr{O}}(\eta,\vec{y}, \vec{y}^{\,\prime}) 
	\bigr]_{\rm g}\bigl| \Omega \bigr\rangle
	=
	\bigl[ \hat{A}_0(\eta,\vec{x}) , \hat{K}_2^\dag(\eta,\vec{y}^{\,\prime}) \bigr]
	\bigl[ \hat{K}_2(\eta,\vec{x}^{\,\prime}) , \hat{A}_0(\eta,\vec{y}) \bigr]
	\neq 0 \, ,
\end{equation}
which is problematic.
It would not be reasonable to conclude that quantum mechanics prevents this object from being a trivial observable,
on account of its correlators not vanishing.
That would imply a significant reduction in the number of observables in the quantized theory, with respect
to the classical theory, and would provide a way to measure a quantum violation of classical first-class constraints. 
The resolution is to require that the product of operators associated with the observable
has to first be properly ordered in the gauge sense,
\begin{align}
\MoveEqLeft[3]
\bigl[ \hat{\mathscr{O}}(\eta,\vec{x},\vec{x}^{\,\prime}) 
	\hat{\mathscr{O}}(\eta, \vec{y}, \vec{y}^{\,\prime})  \bigr]_{\rm g}
\label{OOg}
\\
	={}&
	\hat{K}_2^\dag(\eta,\vec{x}^{\,\prime}) 
	\hat{K}_2^\dag(\eta,\vec{y}^{\,\prime}) 
	\hat{A}_0(\eta,\vec{x})
	\hat{A}_0(\eta,\vec{y})
	+
	\hat{K}_2^\dag(\eta,\vec{x}^{\,\prime}) \hat{A}_0(\eta,\vec{x})
	\hat{A}_0(\eta,\vec{y}) \hat{K}_2(\eta,\vec{y}^{\,\prime} )
\nonumber \\
&
	+
	\hat{K}_2^\dag(\eta,\vec{y}^{\,\prime}) 
	\hat{A}_0(\eta,\vec{x}) 
	\hat{A}_0(\eta,\vec{y})
	\hat{K}_2(\eta,\vec{x}^{\,\prime})
	+
	\hat{A}_0(\eta,\vec{x}) 
	\hat{A}_0(\eta,\vec{y}) 
	\hat{K}_2(\eta,\vec{x}^{\,\prime})
	\hat{K}_2(\eta,\vec{y}^{\,\prime})
	\, ,
\nonumber 
\end{align}
and only then should the expectation value be taken.
This way the two-point function also vanishes,
\begin{equation}
\bigl\langle \Omega \bigr| \bigl[ \hat{\mathscr{O}}(\eta,\vec{x},\vec{x}^{\,\prime}) 
	\hat{\mathscr{O}}(\eta, \vec{y}, \vec{y}^{\,\prime}) \bigr]_{\rm g} \bigl| \Omega \bigr\rangle
	=
	0 \, .
\end{equation}
The operator in~(\ref{OOg}) can also be expressed in terms of the Weyl-ordered products
and compensating commutators, analogous to~(\ref{Og to W}).
The extension of this prescription to higher~$n$-point functions should be straightforward.

\section{Field operator dynamics}
\label{sec: Field operator dynamics}

The dynamics of the linear quantized theory is completely accounted for by the field 
operators. In this section we consider the dynamics of photon field operators
in comoving momentum space, and we
express the solutions in terms of a few scalar mode functions
introduced in Sec.~\ref{subsec: Scalar mode functions}. Canonical commutation relations
fix the normalization of these scalar mode functions, and imply the commutation relations for
time-independent momentum space operators. The section concludes by computing the 
non-Hermitian subsidiary constraint operator, and discussing the freedom in how it
is chosen.

\subsection{Field operators in momentum space}
\label{subsec: Field operators in momentum space}

The FLRW spacetime is homogeneous and isotropic, and
the analysis of dynamics considerably simplifies by working in comoving momentum space. 
It is first advantageous to decompose the spatial components of the vector potential,
\begin{equation}
\hat{A}_i = \hat{A}_i^{\scr T} + \hat{A}_i^{\scr L} \, ,
\qquad \qquad
\hat{\Pi}_i = \hat{\Pi}_i^{\scr T} + \hat{\Pi}_i^{\scr L} \, ,
\label{Ai Pi trans long}
\end{equation}
into its transverse and longitudinal parts,
\begin{equation}
\hat{A}_i^{\scr T} = \mathbb{P}^{\scr T}_{ij} \hat{A}_j \, ,
\qquad
\hat{\Pi}_i^{\scr T} = \mathbb{P}^{\scr T}_{ij} \hat{\Pi}_j \, ,
\qquad
\hat{A}_i^{\scr L} = \mathbb{P}^{\scr L}_{ij} \hat{A}_j \, ,
\qquad
\hat{\Pi}_i^{\scr L} = \mathbb{P}^{\scr L}_{ij} \hat{\Pi}_j \, ,
\end{equation}
defined in terms of the transverse and longitudinal projection operators,
\begin{equation}
\mathbb{P}^{\scr T}_{ij} = \delta_{ij} - \frac{\partial_i \partial_j}{ \nabla^2 } \, ,
\qquad \qquad
\mathbb{P}^{\scr L}_{ij} = \frac{\partial_i \partial_j}{ \nabla^2 } \, ,
\label{projectors}
\end{equation}
that are orthogonal,~$\mathbb{P}_{ij}^{\scr T} \mathbb{P}_{jk}^{\scr L} \!=\! \mathbb{P}_{ij}^{\scr L} \mathbb{P}_{jk}^{\scr T} \!=\!0$, and idempotent,~$\mathbb{P}_{ij}^{\scr T} \mathbb{P}_{jk}^{\scr T} \!=\! \mathbb{P}_{ik}^{\scr T}$,~$\mathbb{P}_{ij}^{\scr L} \mathbb{P}_{jk}^{\scr L} \!=\! \mathbb{P}_{ik}^{\scr L}$.
The conveniently rescaled spatial Fourier transforms of such 
decomposed field operators are
\begin{subequations}
\begin{align}
\hat{A}_0(\eta,\vec{x})
	={}&
	a^{-\frac{D-2-2\zeta}{2} } \!
	\int\! \frac{ d^{D-1}k }{ (2\pi)^{\frac{D-1}{2}} } \,
	e^{i \vec{k} \cdot \vec{x} }
	\, \hat{\mathcal{A}}_0(\eta,\vec{k}) \, ,
\\
\hat{\Pi}_0(\eta,\vec{x})
	={}&
	a^{\frac{D-2-2\zeta}{2} } \!
	\int\! \frac{ d^{D-1}k }{ (2\pi)^{\frac{D-1}{2}} } \,
	e^{i \vec{k} \cdot \vec{x} }
	\, \hat{\pi}_{0}(\eta,\vec{k}) \, ,
\\
\hat{A}_i^{\scr L}(\eta,\vec{x})
	={}&
	a^{-\frac{D-2-2\zeta}{2} } \!
	\int\! \frac{ d^{D-1}k }{ (2\pi)^{\frac{D-1}{2}} } \,
	e^{i \vec{k} \cdot \vec{x} }
	\frac{(-i ) k_i}{k}
	\, \hat{\mathcal{A}}_{\scr L}(\eta,\vec{k}) \, ,
\\
\hat{\Pi}_i^{\scr L}(\eta,\vec{x})
	={}&
	a^{\frac{D-2-2\zeta}{2} } \!
	\int\! \frac{ d^{D-1}k }{ (2\pi)^{\frac{D-1}{2}} } \,
	e^{i \vec{k} \cdot \vec{x} }
	\frac{(-i ) k_i}{k}
	\, \hat{\pi}_{\scr L}(\eta,\vec{k}) \, ,
\\
\hat{A}_i^{\scr T}(\eta,\vec{x})
	={}&
	a^{- \frac{D-4}{2} } \!
	\int\! \frac{ d^{D-1}k }{ (2\pi)^{\frac{D-1}{2}} } \,
	e^{i \vec{k} \cdot \vec{x} }
	\sum_{\sigma=1}^{D-2} \varepsilon_i(\sigma,\vec{k})
	\, \hat{\mathcal{A}}_{{\scr T},\sigma}(\eta,\vec{k}) 
	\, ,
\\
\hat{\Pi}_i^{\scr T}(\eta,\vec{x})
	={}&
	a^{\frac{D-4}{2} } \!
	\int\! \frac{ d^{D-1}k }{ (2\pi)^{\frac{D-1}{2}} } \,
	e^{i \vec{k} \cdot \vec{x} }
	\sum_{\sigma=1}^{D-2} \varepsilon_i(\sigma,\vec{k})
	\, \hat{\pi}_{{\scr T},\sigma}(\eta,\vec{k}) \, ,
\end{align}
\label{Fouriers}%
\end{subequations}
where  the momentum space Hermitian operators behave under conjugation 
as~$\hat{\mathcal{O}}^\dag(\vec{k}) \!=\! \hat{\mathcal{O}}(-\vec{k})$.
Here we introduced transverse polarization tensors with the following properties:
\begin{subequations}
\begin{align}
&
k_i \, \varepsilon_i(\sigma, \vec{k}) = 0 \, ,
&
\quad
\varepsilon_i^* (\sigma,\vec{k})
	= \varepsilon_i(\sigma, - \vec{k}) \, ,
	\qquad \qquad \quad
\\
&
\varepsilon_i^*(\sigma,\vec{k}) \, \varepsilon_i(\sigma', \vec{k})
	= \delta_{\sigma \sigma'} \, ,
\quad
&
\sum_{\sigma=1}^{D-2} \varepsilon_i^*(\sigma, \vec{k}) \varepsilon_j(\sigma, \vec{k})
	= \delta_{ij} - \frac{ k_i k_j }{ k^2 }
	\, .
\end{align}
\end{subequations}
The canonical commutators of the momentum space field operators are now
\begin{subequations}
\begin{align}
&
\bigl[ \hat{\mathcal{A}}_0(\eta,\vec{k} ) , \hat{\pi}_0(\eta,\vec{k}^{\,\prime} ) \bigr]
	= \bigl[ \hat{\mathcal{A}}_{\scr L}(\eta,\vec{k} ) , \hat{\pi}_{\scr L}(\eta,\vec{k}^{\,\prime} ) \bigr]
	= i \delta^{D-1} ( \vec{k} \!+\! \vec{k}^{\,\prime} ) \, ,
\\
&
	\bigl[ \hat{\mathcal{A}}_{{\scr T},\sigma}(\eta,\vec{k} ) , \hat{\pi}_{{\scr T},\sigma'}(\eta,\vec{k}^{\,\prime} ) \bigr]
	= \delta_{\sigma\sigma'} i \delta^{D-1} ( \vec{k} \!+\! \vec{k}^{\,\prime} ) \, ,
\label{transverse momentum commutator}
\end{align}
\label{momentum commutators}%
\end{subequations}
while the momentum space equations of motion for the transverse sector are
\begin{align}
\partial_0 \hat{\mathcal{A}}_{{\scr T},\sigma}
	={}&
	\hat{\pi}_{{\scr T},\sigma} + \frac{1}{2} (D\!-\!4) \mathcal{H} \hat{\mathcal{A}}_{{\scr T},\sigma} \, ,
\label{AT eq}
\\
\partial_0 \hat{\pi}_{{\scr T},\sigma}
	={}&
	- k^2 \hat{\mathcal{A}}_{{\scr T},\sigma}
	- \frac{1}{2} (D\!-\!4) \mathcal{H} \hat{\pi}_{{\scr T},\sigma} \, .
\label{PiT eq}
\end{align}
and the ones for the scalar sector read
\begin{align}
\partial_0 \hat{\mathcal{A}}_{0}
	={}&
	- \xi a^{2-2\zeta} \hat{\pi}_{0}
	+ k \hat{\mathcal{A}}_{\scr L}
	- \frac{1}{2} (D\!-\!2\!-\!2\zeta) \mathcal{H} \hat{\mathcal{A}}_{0} 
	\, ,
\label{A0 eq}
\\
\partial_0 \hat{\pi}_{0}
	={}&
	k \hat{\pi}_{\scr L}
	+ \frac{1}{2} (D\!-\!2\!-\!2\zeta) \mathcal{H} \hat{\pi}_{0} 
	\, ,
\label{Pi0 eq}
\\
\partial_0 \hat{\mathcal{A}}_{\scr L}
	={}&
	a^{2-2\zeta} \hat{\pi}_{\scr L}
	- k \hat{\mathcal{A}}_{0}
	+ \frac{1}{2} (D\!-\!2\!-\!2\zeta) \mathcal{H} \hat{\mathcal{A}}_{\scr L}
	\, ,
\label{AL eq}
\\
\partial_0 \hat{\pi}_{\scr L}
	={}&
	- k \hat{\pi}_{0}
	- \frac{1}{2} (D\!-\!2\!-\!2\zeta) \mathcal{H} \hat{\pi}_{\scr L}
	\, .
\label{PiL eq}
\end{align}
Note that the Fourier transforms in~(\ref{Fouriers}) represent a time-dependent
canonical transformation, so that the momentum space Hamiltonian generating the 
dynamics is
\begin{align}
\hat{{\rm \bf H}}_\star(\eta) 
	={}& 
	\int\! d^{D-1}k 
	\sum_{\sigma=1}^{D-2} 
	\biggl[
	\frac{1}{2} 
	\hat{\pi}_{{\scr T},\sigma}^\dag \, \hat{\pi}_{{\scr T},\sigma}
	+
	\frac{k^2}{2} \hat{\mathcal{A}}_{{\scr T},\sigma}^\dag \, \hat{\mathcal{A}}_{{\scr T},\sigma}
	+
	\frac{(D\!-\!4)}{4} \mathcal{H} 
	\Bigl(
	\hat{\pi}_{{\scr T},\sigma}^\dag \, \hat{\mathcal{A}}_{{\scr T},\sigma}
	+
	\hat{\mathcal{A}}_{{\scr T},\sigma}^\dag \, \hat{\pi}_{{\scr T},\sigma}
	\Bigr)
	\biggr]
\nonumber \\
&	\hspace{-0.5cm}
	+
	\int\! d^{D-1}k \,
	\biggl[
	\frac{a^{2-2\zeta } }{2} \hat{\pi}_{\scr L}^\dag \, \hat{\pi}_{\scr L}
	- 
	\frac{ \xi a^{2-2\zeta } }{2} \hat{\pi}_{0}^\dag \, \hat{\pi}_{0}
	+
	\frac{k}{2} \Bigl(
		\hat{\pi}_{0}^\dag \, \hat{\mathcal{A}}_{\scr L}
		+
		\hat{\mathcal{A}}_{\scr L}^\dag \, \hat{\pi}_{0}
		-
		\hat{\pi}_{\scr L}^\dag \, \hat{\mathcal{A}}_0
		-
		\hat{\mathcal{A}}_0^\dag \, \hat{\pi}_{\scr L}
		\Bigr)
\nonumber \\
&	\hspace{2cm}
	+
	\frac{(D \!-\! 2 \!-\! 2\zeta)}{4} \mathcal{H}
	\Bigl(
		\hat{\pi}_{\scr L}^\dag \, \hat{\mathcal{A}}_{\scr L}
		+
		\hat{\mathcal{A}}_{\scr L}^\dag \, \hat{\pi}_{\scr L}
		-
		\hat{\pi}_{0}^\dag \,\hat{\mathcal{A}}_0
		-
		\hat{\mathcal{A}}_0^\dag \, \hat{\pi}_{0}
		\Bigr)
	\biggr] \, ,
\end{align}
where the arguments of all the fields are~$(\eta,\vec{k})$.
Note that because the system is linear, the operator ordering 
of the Hamiltonian does not matter when
generating the field operator equations of motion. It does matter, however, when 
observables are concerned, as discussed in Sec.~\ref{subsec: Quantum observables}.

\subsection{Subsidiary conditions in momentum space}
\label{subsec: Subsidiary conditions in momentum space}

The subsidiary condition introduced in Sec.~\ref{sec: Subsidiary condition} 
also takes a considerably simpler
form in comoving momentum space. For the Hermitian constraints we define Fourier transforms,
\begin{subequations}
\begin{align}
\hat{\Psi}_1(\eta,\vec{x})
	={}&
	a^{\frac{D-2-2\zeta}{2} } \!
	\int\! \frac{ d^{D-1}k }{ (2\pi)^{\frac{D-1}{2}} } \,
	e^{i \vec{k} \cdot \vec{x} }
	\, \hat{\psi}_{1}(\eta,\vec{k}) \, ,
\\
\hat{\Psi}_2(\eta,\vec{x})
	={}&
	a^{\frac{D-2-2\zeta}{2} } \!
	\int\! \frac{ d^{D-1}k }{ (2\pi)^{\frac{D-1}{2}} } \,
	e^{i \vec{k} \cdot \vec{x} }
	k \, \hat{\psi}_2 (\eta,\vec{k}) \, ,
\end{align}
\end{subequations}
such that the momentum space Hermitian constraints have the same dimensions,
\begin{equation}
\hat{\psi}_1(\eta,\vec{k} ) = \hat{\pi}_0(\eta,\vec{k}) \, ,
\qquad \qquad
\hat{\psi}_2(\eta,\vec{k} ) = \hat{\pi}_{\scr L}(\eta,\vec{k}) \, ,
\end{equation}
and satisfy closed momentum space equations of motion,
\begin{equation}
\partial_0 \hat{\psi}_1 = k \hat{\psi}_2 + \frac{1}{2} (D\!-\!2\!-\!2\zeta) \mathcal{H} \hat{\psi}_1 \, ,
\qquad
\partial_0 \hat{\psi}_2 = - k \hat{\psi}_1 - \frac{1}{2} (D\!-\!2\!-\!2\zeta) \mathcal{H} \hat{\psi}_2 \, .
\label{psi momentum eqs}
\end{equation}
The momentum space non-Hermitian constraint operator is introduced in the same manner,
\begin{equation}
\hat{K}(\vec{x})
	=
	a^{\frac{D-2-2\zeta}{2} } \!
	\int\! \frac{ d^{D-1}k }{ (2\pi)^{\frac{D-1}{2}} } \,
	e^{i \vec{k} \cdot \vec{x} }
	\, \hat{\mathcal{K}} (\vec{k}) \, ,
\end{equation}
which translates into a simple linear combination of the two momentum space Hermitian 
constraint operators,
\begin{equation}
\hat{\mathcal{K}}(\vec{k}) = c_1(\eta,\vec{k}) \hat{\psi}_1(\eta,\vec{k}) 
	+ c_2(\eta, \vec{k}) \hat{\psi}_2(\eta,\vec{k}) \, .
\label{nH constraint}
\end{equation}
The momentum space equivalent of the position space subsidiary condition~(\ref{K position condition}) 
on the space of states now reads
\begin{equation}
\hat{\mathcal{K}}(\vec{k}) \bigl| \Omega \bigr\rangle = 0 \, ,
\qquad \quad
\bigl\langle \Omega \bigr| \hat{\mathcal{K}}^\dag(\vec{k}) = 0 \, ,
\qquad \quad
\forall \vec{k} \, .
\label{subsidiary K momentum}
\end{equation}
The condition of non-Hermiticity of the subsidiary constraint operator,
necessary for consistency with canonical commutation relations,
in momentum space translates into,
\begin{equation}
\hat{\mathcal{K}}^\dag(\vec{k})
	\neq e^{i\gamma(\vec{k}) } \hat{\mathcal{K}}(-\vec{k}) \, ,
\label{non-H condition}
\end{equation}
where~$\gamma(\vec{k})$ is an arbitrary real function reflecting the fact that 
subsidiary conditions~(\ref{subsidiary K momentum}) are defined up to an arbitrary phase.

The conservation of~(\ref{nH constraint}), together with the equations of motion~(\ref{psi momentum eqs})
for the Hermitian constraints, implies equations of motion for the coefficient functions,
\begin{equation}
\partial_0 c_1 = k c_2 - \frac{1}{2} (D\!-\!2\!-\!2\zeta) \mathcal{H} c_1 
	\, ,
\qquad \quad
\partial_0 c_2 = - k c_1+ \frac{1}{2} (D\!-\!2\!-\!2\zeta) \mathcal{H} c_2 
	\, .
\label{c equations}
\end{equation}
The decomposition of the Hermitian constraints in terms of the non-Hermitian ones~(\ref{Hermitian decomposition})
in momentum space now reads
\begin{equation}
\hat{\psi}_1(\eta,\vec{k}) = \hat{\mathcal{K}}_1^\dag(\eta,-\vec{k}) + \hat{\mathcal{K}}_1(\eta,\vec{k}) \, ,
\qquad \quad
\hat{\psi}_2(\eta,\vec{k}) = \hat{\mathcal{K}}_2^\dag(\eta,-\vec{k}) + \hat{\mathcal{K}}_2(\eta,\vec{k}) \, ,
\label{momentum non-H decomposition}
\end{equation}
where
\begin{subequations}
\begin{align}
\hat{K}_1(\eta,\vec{x}) 
	={}& 
	a^{\frac{D-2-2\zeta}{2}}
	\int\! \frac{ d^{D-1}k }{ (2\pi)^{\frac{D-1}{2}} } \,
	e^{i \vec{k} \cdot \vec{x} }
	\, \hat{\mathcal{K}}_{1}(\eta,\vec{k}) \, ,
\\
\hat{K}_2(\eta,\vec{x}) 
	={}& 
	a^{\frac{D-2-2\zeta}{2}}
	\int\! \frac{ d^{D-1}k }{ (2\pi)^{\frac{D-1}{2}} } \,
	e^{i \vec{k} \cdot \vec{x} }
	\,k \, \hat{\mathcal{K}}_{2}(\eta,\vec{k}) \, .
\end{align}
\label{Kfours}%
\end{subequations}
%

\subsection{Solving for dynamics}
\label{subsec: Solving for dynamics}

Solving for dynamics means expressing the time-dependent field operators in the 
Heisenberg picture in terms of initial conditions given at some~$\eta_0$.
This is what the usual solving for operators in terms of creation/annihilation 
operators is, which can also be seen as expressing field operators in the Heisenberg
picture in terms of ones in the Schr{\"o}dinger picture. In this section we express the
solutions of the field operators in terms of the scalar mode functions satisfying 
mode equations, with solutions dependent on the specific FLRW background.

\subsubsection{Transverse sector}

The two transverse sector equations of motion~(\ref{AT eq}) and~(\ref{PiT eq}) combine into
a single second order one,
\begin{align}
\biggl[ \partial_0^2 + k^2
	- \Bigl( \lambda_{\scr T}^2 \!-\! \frac{1}{4} \Bigr) (1\!-\!\epsilon)^2 \mathcal{H}^2 \biggr] 
	\hat{\pi}_{{\scr T},\sigma}
	={}& 0 \, ,
\label{transverse second}
\\
\hat{\mathcal{A}}_{{\scr T},\sigma}
	={}&
	- \frac{1}{k^2}
	\biggl[ \partial_0 + \Bigl( \lambda_{\scr T} \!+\! \frac{1}{2} \Bigr) (1\!-\!\epsilon) \mathcal{H} \biggr]
	\hat{\pi}_{{\scr T},\sigma} \, ,
\end{align}
where we introduce
\begin{equation}
\lambda_{\scr T} = \frac{D\!-\!5\!+\!\epsilon}{2(1\!-\!\epsilon)} \, .
\end{equation}
The second order equation~(\ref{transverse second}) is just the
scalar mode equation~(\ref{scalar mode eq}) with~$\lambda\!\to\! \lambda_{\scr T}$.
Furthermore, we have that~$(\lambda_{\scr T}\!+\! \frac{1}{2})(1\!-\!\epsilon) \!=\! (D\!-\!4)/2$ is time-independent,
so that recurrence relations~(\ref{recurrences})  are applicable. Therefore, we can
write the solutions as
\begin{align}
\hat{\pi}_{{\scr T},\sigma} (\eta,\vec{k})
	={}&
	- i k \, \mathcal{U}_{\lambda_{\scr T}}(\eta,k) \, \hat{b}_{\scr T}(\sigma,\vec{k})
	+
	i k \, \mathcal{U}_{\lambda_{\scr T}}^*(\eta,k) \, \hat{b}_{\scr T}^\dag(\sigma,-\vec{k})
	\, ,
\label{PiT solution}
\\
\hat{\mathcal{A}}_{{\scr T},\sigma} (\eta,\vec{k})
	={}&
	\mathcal{U}_{\lambda_{\scr T}+1}(\eta,k) \, \hat{b}_{\scr T}(\sigma,\vec{k})
	+
	\mathcal{U}_{\lambda_{\scr T}+1}^*(\eta,k) \, \hat{b}_{\scr T}^\dag(\sigma,-\vec{k})
	\, .
\label{AT solution}
\end{align}
It follows now from the canonical commutation relations~(\ref{transverse momentum commutator})
and the Wronskian~(\ref{Wronskian}) that the initial condition
operators satisfy creation/annihilation commutation relations,
\begin{equation}
\bigl[ \hat{b}_{\scr T}(\sigma, \vec{k}) , \hat{b}_{\scr T}^\dag(\sigma', \vec{k}^{\,\prime}) \bigr]
	= 
	\delta_{\sigma \sigma'} \,
	\delta^{D-1}(\vec{k} \!-\! \vec{k}^{\,\prime} ) \, .
\end{equation}
Explicit solutions for the transverse sector mode functions depend on the 
particular FLRW background only, and not on the gauge-fixing parameters.
This reflects the fact that the transverse polarizations are the physical propagating degrees
of freedom of the photon in spatially flat cosmological spaces.

\subsubsection{Scalar sector}

In the scalar sector the two equations
for canonical momenta~(\ref{Pi0 eq}) and~(\ref{PiL eq})
decouple from the rest.
They combine into a second order equation,
\begin{align}
\biggl[ \partial_0^2 + k^2 
	- \Bigl( \lambda^2 \!-\! \frac{1}{4} \Bigr) (1\!-\!\epsilon)^2 \mathcal{H}^2 \biggr] \hat{\pi}_{\scr L}
	={}& 0 \, ,
\\
\hat{\pi}_0 ={}&
	- \frac{1}{k} \biggl[ \partial_0 
	+ \Bigl( \lambda \!+\! \frac{1}{2} \Bigr)
		(1\!-\!\epsilon) \mathcal{H} \biggr] \hat{\pi}_{\scr L}
		\, ,
\end{align}
taking the form of the scalar mode equation,
where the parameter
\begin{equation}
\lambda = \frac{ D \!-\! 3 \!+\!\epsilon \!-\! 2\zeta }{ 2(1\!-\!\epsilon) }
\label{lambda def}
\end{equation}
satisfies the relation~(\ref{time relation}). Therefore, 
according to~(\ref{scalar mode eq}),~(\ref{general solution}), and~(\ref{recurrences}),
the solutions are given in terms of scalar mode functions,
\begin{align}
\hat{\pi}_{\scr L}(\eta,\vec{k}) ={}&
	k \,\mathcal{U}_\lambda(\eta,k) \, \hat{b}_{\scr P}(\vec{k})
	+ k \, \mathcal{U}_\lambda^*(\eta,k) \, \hat{b}_{\scr P}^\dag(-\vec{k}) 
	\, ,
\\
\hat{\pi}_{0}(\eta,\vec{k}) ={}&
	ik \, \mathcal{U}_{\lambda+1}(\eta,k) \, \hat{b}_{\scr P}(\vec{k})
	- ik \, \mathcal{U}_{\lambda+1}^*(\eta,k) \, \hat{b}_{\scr P}^\dag(-\vec{k}) 
	\, .
\end{align}
These solutions now source the two
remaining scalar sector equations~(\ref{A0 eq}) and~(\ref{AL eq}), 
which again combine into 
a single second order one,
\begin{align}
&
\biggl[ \partial_0^2 + k^2 
	- \Bigl( \lambda^2 \!-\! \frac{1}{4} \Bigr) (1\!-\!\epsilon)^2 \mathcal{H}^2 \biggr] 
		\hat{\mathcal{A}}_0
	=
	a^{2-2\zeta} \biggl[
	-2\xi(1\!-\!\zeta) \mathcal{H} \hat{\pi}_{0}
	+  (1 \!-\! \xi ) k \hat{\pi}_{\scr L}
	\biggr] \, ,
\label{A0 mode eq}
\\
&
\hspace{4cm}
\hat{\mathcal{A}}_{\scr L} =
	\frac{1}{k} \biggl[
		\partial_0 + \Bigl( \lambda \!+\! \frac{1}{2} \Bigr) (1\!-\!\epsilon) \mathcal{H}
		\biggr] \hat{\mathcal{A}}_0
		+ \frac{\xi a^{2-2\zeta}}{k} \hat{\pi}_0
	\, .
\label{AL mode eq}
\end{align}
The source for the second order equation above already suggests what is likely the simplest choice of 
gauge-fixing parameters
--- $\xi \!=\! 1$ and~$\zeta \!=\! 1$ --- which turns it into a homogeneous one.
In the de Sitter space limit $\epsilon\!=\!0$, this corresponds to 
the simple noncovariant gauge due to Woodard~\cite{Woodard:2004ut}.
Equations~(\ref{A0 mode eq}) and~(\ref{AL mode eq}) are solved by
\begin{align}
\hat{\mathcal{A}}_0 (\eta,\vec{k})
	={}&
	\mathcal{U}_\lambda(\eta,k) \, \hat{b}_{\scr H}(\vec{k})
	+ \mathcal{U}_\lambda^*(\eta,k) \, \hat{b}_{\scr H}^\dag(-\vec{k})
\nonumber \\
&	\hspace{4cm}
	+ v_0(\eta,k) \, \hat{b}_{\scr P}(\vec{k})
	+ v_0^*(\eta,k) \, \hat{b}_{\scr P}^\dag(-\vec{k}) \, ,
\\
\hat{\mathcal{A}}_{\scr L} (\eta,\vec{k})
	={}&
	- i \, \mathcal{U}_{\lambda+1}(\eta,k) \, \hat{b}_{\scr H}(\vec{k})
	+ i \,  \mathcal{U}_{\lambda+1}^*(\eta,k) \, \hat{b}_{\scr H}^\dag(-\vec{k})
\nonumber \\
&	\hspace{4cm}
	- i v_{\scr L} (\eta,k) \, \hat{b}_{\scr P}(\vec{k})
	+ i v_{\scr L}^*(\eta,k) \, \hat{b}_{\scr P}^\dag(-\vec{k}) \, ,
\end{align}
where the homogeneous parts solve the scalar mode equation~(\ref{scalar mode eq}),
while the particular mode functions~$v_0$ and~$v_{\scr L}$ satisfy sourced mode equations,
\begin{align}
&
\biggl[ \partial_0^2 + k^2
	- \Bigl( \lambda^2 \!-\! \frac{1}{4} \Bigr) (1\!-\!\epsilon)^2 \mathcal{H}^2 \biggr] v_0
	=
	a^{2-2\zeta} \biggl[
	- 2 i \xi ( 1 \!-\! \zeta ) k\mathcal{H} \, \mathcal{U}_{\lambda+1} 
	+ (1\!-\!\xi) k^2 \mathcal{U}_\lambda
	\biggr] \, ,
\\
&	\hspace{4cm}
v_{\scr L} =
	\frac{i}{k} \biggl[
		\partial_0 + \Bigl( \lambda \!+\! \frac{1}{2} \Bigr) (1\!-\!\epsilon) \mathcal{H}
		\biggr] v_0
		- \xi a^{2-2\zeta} \mathcal{U}_{\lambda+1}
	\, .
\end{align}
and we conveniently normalize them to
\begin{equation}
{\rm Re} \Bigl[ v_0(\eta,k) \, \mathcal{U}_{\lambda+1}^*(\eta,k)
	+ v_{\scr L}(\eta,k) \, \mathcal{U}_{\lambda}^*(\eta,k) \Bigr]
	= 0
	\, .
\end{equation}
This fixes the commutation relations between the time-independent operators,
the only nonvanishing ones being
\begin{equation}
\bigl[ \hat{b}_{\scr P}(\vec{k}) , \hat{b}_{\scr H}^\dag(\vec{k}^{\,\prime}) \bigr]
=
\bigl[ \hat{b}_{\scr H}(\vec{k}) , \hat{b}_{\scr P}^\dag(\vec{k}^{\,\prime}) \bigr]
	= - \delta^{D-1}( \vec{k} \!-\! \vec{k}^{\,\prime} ) 
	\, .
\label{bPbH commutators}
\end{equation}
These are not the canonical commutation relations for the creation and annihilation operators
that one is accustomed to working with. Nonetheless, they are perfectly valid solutions.
In fact, a simple non-Bogolyubov transformation,
\begin{equation}
\hat{b}_1(\vec{k}) = \frac{1}{\sqrt{2}} \Bigl( \hat{b}_{\scr H}^\dag(-\vec{k}) + \hat{b}_{\scr P}^\dag(-\vec{k}) \Bigr) \, ,
\qquad \qquad
\hat{b}_2(\vec{k}) = \frac{1}{\sqrt{2}} \Bigl( \hat{b}_{\scr H}(\vec{k}) - \hat{b}_{\scr P}(\vec{k}) \Bigr) \, ,
\end{equation}
leads to more familiar creation/annihilation operators with nonvanishing commutators,
\begin{equation}
\bigl[ \hat{b}_1(\vec{k}) , \hat{b}_1^\dag(\vec{k}^{\,\prime}) \bigr] 
	=
\bigl[ \hat{b}_2(\vec{k}) , \hat{b}_2^\dag(\vec{k}^{\,\prime}) \bigr] = \delta^{D-1}(\vec{k} \!-\! \vec{k}^{\,\prime}) \, .
\end{equation}
However, for our purposes it is far more convenient to work with~$\hat{b}_{\scr P}$
and~$\hat{b}_{\scr H}$ directly, as they readily translate to the subsidiary condition on
physical states.

\subsubsection{Non-Hermitian subsidiary constraint}
\label{subsubsec: Non-Hermitian subsidiary constraint}

The conservation of the non-Hermitian subsidiary constraint operator~(\ref{nH constraint})
implies two Eqs.~(\ref{c equations}) that combine into a single second order one,
\begin{align}
\biggl[ \partial_0^2 + k^2 
	- \Bigl( \lambda^2 \!-\! \frac{1}{4} \Bigr) (1\!-\!\epsilon)^2 \mathcal{H}^2 \biggr] c_1 ={}&
	0
	 \, ,
\\
c_2 ={}&
	\frac{1}{k} \biggl[ \partial_0 + \Bigl( \lambda \!+\! \frac{1}{2} \Bigr) 
		(1\!-\!\epsilon) \mathcal{H} \biggr] c_1 
		\, ,
\end{align}
with~$\lambda$ defined in~(\ref{lambda def}).
Again we recognize the scalar mode equation~(\ref{scalar mode eq})
and apply the recurrence relation~(\ref{recurrences}) to write the general solutions
in a convenient form:
\begin{align}
c_1(\eta,\vec{k}) ={}&
	i \beta(-\vec{k}) \, \mathcal{U}_\lambda(\eta,k) 
		- i \alpha(\vec{k}) \, \mathcal{U}_\lambda^*(\eta,k)
	\, ,
\\
c_2(\eta,\vec{k}) ={}&
	\beta(-\vec{k}) \, \mathcal{U}_{\lambda+1}(\eta,k) 
		+ \alpha(\vec{k}) \, \mathcal{U}_{\lambda+1}^*(\eta,k)
	\, ,
\end{align}
where~$\alpha(\vec{k})$ and~$\beta(\vec{k})$ are free coefficients.
Upon using the Wronskian~(\ref{Wronskian})
the non-Hermitian constraint evaluates to
\begin{equation}
\hat{\mathcal{K}}(\vec{k})
	=
	 \alpha(\vec{k}) \, \hat{b}_{\scr P}(\vec{k})
	+ \beta(-\vec{k}) \, \hat{b}_{\scr P}^\dag(-\vec{k}) \, ,
\label{K solution}
\end{equation}
and the condition of non-Hermiticity~(\ref{non-H condition})
now translates into the condition on the coefficients,
\begin{equation}
\biggl| \frac{\alpha(\vec{k})}{\beta(\vec{k})} \biggr| \neq 1 \, .
\label{non-hermiticity condition}
\end{equation}
The way that the free coefficients appear in~(\ref{K solution})
is reminiscent of Bogolyubov coefficients, and ultimately they have such an interpretation.
Since the overall normalization 
of~$\hat{\mathcal{K}}$ is immaterial we may 
parametrize it conveniently
as~\footnote{
This
parametrization technically covers just half of the parameter space. The other half is covered by 
interchanging the roles of~$\hat{b}_{\scr P}(\vec{k})$ and~$\hat{b}_{\scr P}^\dag(-\vec{k}).$
Even though there should be no obstructions to this choice it turns out to be inconsistent with manifest
Poincar\'e symmetries in flat space, and we do not consider it.}
\begin{equation}
\hat{\mathcal{K}}(\vec{k})
	= \mathcal{N}(\vec{k}) \,
	e^{i \theta(\vec{k})} \Bigl( e^{-i \varphi(\vec{k})} {\rm ch}[ \rho(\vec{k}) ] \, \hat{b}_{\scr P}(\vec{k}) 
		+ e^{ i \varphi(-\vec{k})} {\rm sh}[ \rho(-\vec{k}) ] \, \hat{b}_{\scr P}^\dag(-\vec{k}) \Bigr)
		\, ,
\label{K parametrized}
\end{equation}
where we introduced a normalization coefficient,
\begin{equation}
\mathcal{N}(\vec{k}) = \Bigl( {\rm ch}[\rho(\vec{k})] \, {\rm ch}[\rho(-\vec{k})]
	- {\rm sh}[\rho(\vec{k})] \, {\rm sh}[\rho(-\vec{k})] \Bigr)^{\!\! - \frac{1}{2}}
	\! =
	\Bigl( {\rm ch} \bigl[ \rho(\vec{k}) \!-\! \rho(- \vec{k}) \bigr] \Bigr)^{\!\! - \frac{1}{2} }
	\, ,
\end{equation}
and where~$\theta(\vec{k})$, $\varphi(\vec{k})$, and~$\rho(\vec{k})$
are arbitrary real functions.
Fixing these functions is a matter of convenience, whether it is respecting some symmetry or some other requirement.
Since~$\hat{\mathcal{K}}(\vec{k})$ will annihilate the ket state, it is convenient to employ it
in computations, instead of using~$\hat{b}_{\scr P}(\vec{k})$. It is likewise
advantageous to introduce another non-Hermitian operator associated 
with~$\hat{b}_{\scr H}(\vec{k})$,
\begin{equation}
\hat{\mathcal{B}}(\vec{k})
	= \mathcal{N}(\vec{k}) \,
	e^{ i \theta(\vec{k})} \Bigl( e^{ - i \varphi(\vec{k})} {\rm ch}[ \rho(-\vec{k}) ] \, \hat{b}_{\scr H}(\vec{k}) 
		+ e^{ i \varphi(-\vec{k})} {\rm sh}[ \rho(\vec{k}) ] \, \hat{b}_{\scr H}^\dag(-\vec{k}) \Bigr)
		\, .
\label{B parametrized}
\end{equation}
that preserves the form of nonvanishing commutators~(\ref{bPbH commutators}),
\begin{equation}
\bigl[ \hat{\mathcal{K}}(\vec{k}) , \hat{\mathcal{B}}^\dag(\vec{k}^{\, \prime}) \bigr]
	=
	\bigl[ \hat{\mathcal{B}}(\vec{k}) , \hat{\mathcal{K}}^\dag(\vec{k}^{\, \prime}) \bigr]
	=
	- \delta^{D-1} ( \vec{k} \!-\! \vec{k}^{\,\prime} )
	\, .
\label{KB commutator}
\end{equation}
In this sense~(\ref{K parametrized})--(\ref{B parametrized}) can be seen as a Bogolyubov transformation
preserving the noncanonical commutation relations~(\ref{bPbH commutators}).

\medskip

We can now also evaluate the parts of the non-Hermitian decomposition~(\ref{momentum non-H decomposition})
in terms of the scalar mode functions,
that will prove useful later,
\begin{subequations}
\begin{align}
\hat{\mathcal{K}}_1(\eta,\vec{k}) \!={}&\!
	i k \mathcal{N}(\vec{k}) e^{-i\theta(\vec{k})} \!
	\biggl[
	e^{i\varphi(\vec{k})} {\rm ch}[\rho(-\vec{k})] \mathcal{U}_{\lambda+1}(\eta,k)
	\!+\!
	e^{-i\varphi(-\vec{k})} {\rm sh}[\rho(\vec{k})] \mathcal{U}_{\lambda+1}^*(\eta,k)
	\biggr]
	\hat{\mathcal{K}}(\vec{k})
	\, ,
\\
\hat{\mathcal{K}}_2(\eta,\vec{k}) ={}&
	k \mathcal{N}(\vec{k}) e^{-i\theta(\vec{k})}
	\biggl[
	e^{i\varphi(\vec{k})} {\rm ch}[\rho(-\vec{k})] \mathcal{U}_{\lambda}(\eta,k)
	-
	e^{-i\varphi(-\vec{k})} {\rm sh}[\rho(\vec{k})] \mathcal{U}_{\lambda}^*(\eta,k)
	\biggr]
	\hat{\mathcal{K}}(\vec{k})
	\, .
\end{align}
\label{K1K2}%
\end{subequations}
%

\section{Constructing the space of states}
\label{sec: Constructing space of states}

The preceding section considered the quantization of the dynamics of field operators,
and of the subsidiary non-Hermitian 
constraint operator. To complete the quantization we need to
construct a space of states on which the field operators act. This cannot be the usual Fock
space due to the subsidiary condition~(\ref{subsidiary K momentum}) that forces upon
us an indefinite inner product space. 
The construction of the space of states in quantized theories is typically intricately connected 
to the symmetries of the system. 
Here we discuss two concepts of symmetries arising in
multiplier gauges: {\it physical symmetries} that 
are symmetries of the gauge-invariant 
action~(\ref{invariant action}),
and {\it gauge-fixed symmetries} that are symmetries of 
the gauge-fixed action~(\ref{fixed action}).
The former are actual symmetries of the system, and characterize physical properties of the state,
while the latter are symmetries of the
gauge-fixed dynamics and are a matter of choice. Even though in the case at hand the
gauge-fixed symmetries coincide with the physical symmetries, as both actions~(\ref{invariant action})
and~(\ref{fixed action}) are
invariant under spatial Euclidean transformations, in general this need not be the case.
For example, in the de Sitter space limit ($\epsilon\!=\!0$) the gauge-invariant action would 
be invariant under the maximal number of isometries, while the gauge-fixed action for~$\zeta\!\neq\!0$
would be invariant under Euclidean spatial transformations only. Thus we would be able to define 
a state respecting physical de Sitter symmetries, but the gauge-fixed dynamics could not be
made de Sitter invariant. 
For such a state correlators of gauge-independent operators would exhibit physical symmetries,
despite the fact that the correlators of gauge-dependent quantities would not.
This is why understanding the distinction between the two is important.
The physical symmetries will influence the construction of the transverse sector
of the space of states, while the gauge-fixed symmetries will dictate the construction 
of the scalar sector.

\subsection{FLRW symmetries}
\label{subsec: FLRW symmetries}

Flat FLRW spacetimes have~$\frac{1}{2}D(D\!-\!1)$ isometries of the~$(D\!-\!1)$-dimensional Euclidean
spaces that make the equal-time spatial slices. They consist of~$(D\!-\!1)$ spatial translations,
\begin{equation}
\eta \longrightarrow \eta \, ,
\qquad\qquad
x_i \longrightarrow x_i + \alpha_i \, ,
\end{equation}
and of~$\frac{1}{2}(D\!-\!1)(D\!-\!2)$ spatial rotations, whose infinitesimal form is
\begin{equation}
\eta \longrightarrow \eta \, ,
\qquad \qquad
x_i \longrightarrow x_i + 2 \omega_{ij} x_j \, ,
\qquad \qquad(\omega_{ij} \!=\! - \omega_{ji})
\, .
\end{equation}
Both the gauge-invariant and the gauge-fixed photon actions,~(\ref{invariant action}) 
and~(\ref{fixed action}),
are invariant under infinitesimal active transformations of the vector potential, 
associated with spatial translations,
\begin{equation}
A_\mu(x) \longrightarrow A_\mu(x) - \alpha_i \partial_i A_\mu(x) \, ,
\label{active translation}
\end{equation}
and with spatial rotations,
\begin{equation}
A_\mu(x) \longrightarrow A_\mu(x) + 2 \omega_{ij} x_i \partial_j A_\mu (x)
	+ 2 \delta_\mu^i \omega_{ij} A_j(x) \, .
\label{active rotation}
\end{equation}
%

\subsection{Physical symmetries}
\label{subsec: Physical symmetries}

Even though the active transformations~(\ref{active translation}) and~(\ref{active rotation})
are symmetry transformations of the gauge-invariant action~(\ref{invariant action}),
they are ambiguous on the account of gauge transformations that carry no physical meaning.
This means we can combine~(\ref{active translation}) and~(\ref{active rotation})
with a gauge transformation, and change their form without affecting the physical content.
The most convenient choice fixing the ambiguity is requiring that the generators of these
transformations take a gauge-invariant form themselves. This is accomplished by 
modifying~(\ref{active translation}) and~(\ref{active rotation}) by a gauge transformation to read,
respectively,
\begin{equation}
A_\mu(x) \longrightarrow A_\mu(x) - \alpha_i F_{i\mu}(x) \, ,
\qquad \qquad
A_\mu(x) \longrightarrow A_\mu(x)+ 2 \omega_{ij} x_i F_{j\mu} (x) \, .
\end{equation}
Thus, the conserved Noether charges associated with the two symmetry transformations 
of the gauge-invariant action~(\ref{invariant action}) are, respectively,
the total linear momentum and total angular momentum,
\begin{equation}
P_i =
	\int\! d^{D-1}x \, \Bigl( - F_{ij} \Pi_j \Bigr) \, ,
\qquad \qquad
M_{ij} = \int\! d^{D-1}x \, \Bigl( 2 x_{[i} F_{j]k} \Pi_k \Bigr) \, .
\end{equation}
They satisfy~$E(D\!-\!1)$ algebra on-shell,
\begin{equation}
\bigl\{ P_i , P_j \bigr\} \approx 0 
	\, ,
\qquad
\bigl\{ M_{ij} , P_k \bigr\} \approx 
	2 P_{[i} \delta_{j]k} 
	\, ,
\qquad
\bigl\{ M_{ij} , M_{kl} \bigr\} \approx 
	4 \delta_{i] [k} M_{l] [j}
	\, ,
\label{gauge invariant algebra}
\end{equation}
and serve as generators of corresponding symmetry transformations.
Their structure is more transparent if we write them out in terms of longitudinal and transverse components
of the canonical variables~(\ref{Ai Pi trans long})--(\ref{projectors}), 
and recognize the constraints~(\ref{configuration space constraints}),
\begin{align}
P_i ={}& 
	\int\! d^{D-1}x \, \Bigl( - \Pi_j^{\scr T} \partial_i A_j^{\scr T} 
		- A_i^{\scr T} \Psi_2 \Bigr)
	\, ,
\label{Pi inv}
\\
M_{ij} ={}&
	\int\! d^{D-1}x \, 
	\Bigl(
	2 x_{[i} F_{j]k}^{\scr T} \Pi_k^{\scr T}
	+ 2 x_{[i} A_{j]}^{\scr T} \Psi_2
		\Bigr)
	\, .
\label{Mij inv}
\end{align}
Quantizing these symmetry generators implies promoting fields to field operators,
which necessitates proper operator ordering in order for them to be observables. 
First, the parts of~(\ref{Pi inv}) and~(\ref{Mij inv}) containing constraints
should be ordered according to the prescription outlined in 
Sec.~\ref{subsec: Quantum observables},
\begin{align}
\hat{P}_i ={}&
	\hat{P}_i^{\scr T}
	-
	\int\! d^{D-1}x \, 
	\Bigl(
	\hat{K}_2^\dag \hat{A}_i^{\scr T}
	+ \hat{A}_i^{\scr T} \hat{K}_2 
	\Bigr)
	\, ,
\label{hatPi}
\\
\hat{M}_{ij} ={}&
	\hat{M}_{ij}^{\scr T}
	+
	\int\! d^{D-1}x \, 
	\Bigl(
	2 \hat{K}_2^\dag x_{[i} \hat{A}_{j]}^{\scr T} 
	+
	2 x_{[i} \hat{A}_{j]}^{\scr T} \hat{K}_2
	\Bigr)
	\, ,
\label{hatMij}
\end{align}
Second, the purely transverse parts should be normal-ordered according to
the standard prescription that is best implemented in momentum space where
all the annihilation operators of the transverse sector are put to the right
of all the creation operators.
Using Fourier transforms of field operators~(\ref{Fouriers}),
the solutions of the transverse field operators~(\ref{AT solution}) and~(\ref{PiT solution}),
and the Wronskian~(\ref{Wronskian}) of the mode function,
the normal-ordered purely transverse parts of the operators evaluate to
\begin{align}
\hat{P}_i^{\scr T} ={}&
		\int\! d^{D-1}k \, k_i \, \hat{\mathcal{E}}_j^\dag(\vec{k}) \hat{\mathcal{E}}_j(\vec{k})
		\, ,
\label{PiT}
\\
\hat{M}_{ij}^{\scr T} ={}&
\int\! d^{D-1} k
	\biggl[
	\hat{\mathcal{E}}_k^\dag(\vec{k})
	\biggl( i k_i \frac{\partial}{\partial k_j} \!-\! i k_j \frac{\partial}{\partial k_i} \biggr)
	\hat{\mathcal{E}}_k(\vec{k})
	+
	2 \hat{\mathcal{E}}_{[i}^\dag(\vec{k}) \hat{\mathcal{E}}_{j]}(\vec{k})
	\biggr]
	\, , 
\label{MijT}
\end{align}
with the expressions written compactly using a shorthand notation,
\begin{equation}
\hat{\mathcal{E}}_i(\vec{k})
	=
	\sum_{\sigma=1}^{D-2} \varepsilon_i(\sigma,\vec{k}) \, \hat{b}_{\scr T}(\sigma,\vec{k}) 
	\, ,
\label{B def}
\end{equation}
These generators of physical symmetries commute with the non-Hermitian constraint,
\begin{equation}
\bigl[ \hat{\mathcal{K}}(\vec{k}) , \hat{P}_i \bigr] = 0 \, ,
\qquad \qquad
\bigl[ \hat{\mathcal{K}}(\vec{k}) , \hat{M}_{ij} \bigr] = 0 \, ,
\end{equation}
and preserve the algebra~(\ref{gauge invariant algebra}) at the level of matrix elements,
\begin{align}
&
\bigl\langle \psi \bigr| \bigl[ \hat{P}_i , \hat{P}_j \bigr] \bigl| \psi' \bigr\rangle = 0 
\, ,
\qquad\quad
\bigl\langle \psi \bigr| \bigl[ \hat{M}_{ij} , \hat{P}_k \bigr] \bigl| \psi' \bigr\rangle = 
	i \bigl\langle \psi \bigr| 2 \hat{P}_{[i} \delta_{j]k} \bigl| \psi' \bigr\rangle 
\, ,
\nonumber \\
&	\hspace{2cm}
\bigl\langle \psi \bigr| \bigl[ \hat{M}_{ij} , \hat{M}_{kl} \bigr] \bigl| \psi' \bigr\rangle =
	i \bigl\langle \psi \bigr| 4 \delta_{i] [k} M_{l] [j} \bigl| \psi' \bigr\rangle
	\, .
\label{matrix element algebra}
\end{align}
In fact, it is only the purely transverse parts of~(\ref{hatPi}) and~(\ref{MijT}) 
that contribute to the matrix elements of the algebra.

The dynamics of field operators is given by the gauge-fixed action. As a consequence
the physical symmetry generators~(\ref{hatPi}) and~(\ref{hatMij}) are not time-independent.
That is why it is meaningless to require there exists a state that is an eigenstate 
of these generators in the usual sense. However, the matrix elements of physical
symmetry generators are conserved in time, since only the purely transverse 
part provides a nonvanishing contribution to them.  The fact that this part 
is time-independent can be seen from the solutions given in~(\ref{PiT})
and~(\ref{MijT}). This implies that we can still define a notion of an 
eigenstate~$\bigl| \Omega \bigr\rangle$
of generators by the property that an expectation value of any polynomial 
of generators equals the polynomial of expectation values,
\begin{equation}
\bigl\langle \Omega \bigr| \mathscr{P} \bigl( \hat{P}_i , \hat{M}_{ij} \bigr) \bigl| \Omega \bigr\rangle
	= \mathscr{P} \bigl( \overline{P}_i , \overline{M}_{ij} \bigr)
	\, ,
	\qquad
	\bigl\langle \Omega \bigr| \hat{P}_i \bigl| \Omega \bigr\rangle = \overline{P}_i
	\, ,
	\quad
	\bigl\langle \Omega \bigr| \hat{M}_{ij} \bigl| \Omega \bigr\rangle = \overline{M}_{ij}
	\, .
\label{different eigenvalue condition}
\end{equation}
This condition, together with the algebra~(\ref{matrix element algebra}), 
implies there is only one such state that is a simultaneous eigenstate of both,
and that it has to have vanishing expectation values,
\begin{equation}
\overline{P}_{i}\!=\!0 \, ,\qquad \qquad \overline{M}_{ij} = 0 \, .
\end{equation}
Another thing becomes evident upon closer examination of~(\ref{different eigenvalue condition}) 
--- it is only the transverse sector
that is affected by these conditions, and the full state will be the tensor product between
the transverse sector and the scalar 
sector,~$\bigl| \Omega \bigr\rangle \!=\! \bigl| \Omega_{\scr T} \bigr\rangle \otimes \bigl| \Omega_0 \bigr\rangle$.
The only scalar sector operators appearing in 
the generators are constraints, which are ordered such that terms containing them drop
out from any expectation values:
\begin{equation}
\bigl\langle \Omega \bigr| \mathscr{P} \bigl( \hat{P}_i , \hat{M}_{ij} \bigr) \bigl| \Omega \bigr\rangle
	=
	\bigl\langle \Omega_{\scr T} \bigr| \mathscr{P} \bigl( \hat{P}_i^{\scr T} , \hat{M}_{ij}^{\scr T} \bigr) \bigl| \Omega_{\scr T} \bigr\rangle
	\, .
\end{equation}

Since the transverse sector is unconstrained
its space of states can be constructed as the usual Fock space. 
There must be some annihilation operator~$\hat{c}_{\scr T}(\sigma,\vec{k})$ that annihilates the 
vacuum,
\begin{equation}
\hat{c}_{\scr T}(\sigma,\vec{k}) \bigl| \Omega_{\scr T} \bigr\rangle=0 \, ,
\qquad \quad
\forall \sigma, \vec{k} \, ,
\label{transverse vacuum}
\end{equation}
so that the rest of the Fock space is 
generated by acting with the
associated creation operators~$\hat{c}_{\scr T}^\dag(\sigma,\vec{k})$ on that vacuum. 
The most general choice respecting isotropy and homogeneity is given by the Bogolyubov
transformation,
\begin{equation}
\hat{c}_{\scr T}(\sigma,\vec{k}) = 
	e^{-i \varphi_{\scr T}(k) } {\rm ch}[\rho_{\scr T}(k)] \, \hat{b}_{\scr T}(\sigma,\vec{k})
	+ e^{i \varphi_{\scr T}(k)} {\rm sh}[\rho_{\scr T}(k)] \, \hat{b}_{\scr T}^\dag(\sigma,-\vec{k})
	\, ,
\label{cT}
\end{equation}
where~$\varphi_{\scr T}(k)$ and~$\rho_{\scr T}(k)$ are arbitrary real functions. 
The vacuum defined in~(\ref{transverse vacuum}) is now an eigenstate of the
purely transverse parts of generators~(\ref{PiT}) and~(\ref{MijT}),
\begin{equation}
\hat{P}_i^{\scr T} \bigl| \Omega_{\scr T} \bigr\rangle = 0 \, ,
\qquad \qquad
\hat{M}_{ij}^{\scr T} \bigl| \Omega_{\scr T} \bigr\rangle = 0 \, ,
\end{equation}
with vanishing eigenvalues, and thus corresponds to the state respecting physical 
cosmological symmetries. Note that the procedure of this section has fixed
only the transverse sector of the state, while leaving the scalar sector unfixed.
The scalar sector has to be fixed from different considerations, that the following
section is devoted to.

\subsection{Gauge-fixed symmetries}
\label{subsec: Gauge-fixed symmetries}

The conserved Noether charges associated with spatial translations and spatial rotations that follow from the
gauge-fixed action~(\ref{fixed action}) are, respectively,
\begin{align}
P_i^\star ={}&
	\int\! d^{D-1}x \, \Bigl(
		- \Pi_0 \partial_i A_0
		- \Pi_j \partial_i A_j
		\Bigr)
		\, ,
\\
M_{ij}^\star ={}&
	\int\! d^{D-1}x \, \Bigl(
	2 \Pi_0 x_{[i} \partial_{j]} A_0
	+ 2 \Pi_k x_{[i} \partial_{j]} A_k
	+ 2 \Pi_{[i} A_{j]}
	\Bigr)
	\, .
\end{align}
They are generators of the corresponding symmetry transformations
of the gauge-fixed dynamics,
and they satisfy the $E(D\!-\!1)$ algebra off-shell,
\begin{equation}
\bigl\{ P_i^\star , P_j^\star \bigr\} = 0 
	\, ,
\qquad
\bigl\{ M_{ij}^\star , P_k^\star \bigr\} =
	2 P_{[i}^\star \delta_{j]k}
	\, ,
\qquad
\bigl\{ M_{ij}^\star , M_{kl}^\star \bigr\} =
	4 \delta_{i] [k} M_{l] [j}^\star
	\, ,
\label{gauge fixed algebra}
\end{equation}
The structure of these charges and their quantization 
is more transparent when written in terms of transverse
and longitudinal components of the canonical fields,
\begin{align}
P_i^\star ={}&
	\int\! d^{D-1}x \, \biggl[
		- \Pi_j^{\scr T} \partial_i A_j^{\scr T}
		+ \Bigl( \Psi_2 A_i^{\scr L}
			- \Psi_1 \partial_i A_0 \Bigr)
		\biggr]
		\, ,
\\
M_{ij}^\star ={}&
	\int\! d^{D-1}x \, \biggl[
		2 x_{[i} F_{j]k}^{\scr T} \Pi_k^{\scr T}
		+ 2 \Bigl( \Psi_1 x_{[i} \partial_{j]} A_0
			- \Psi_2 x_{[i} A_{j]}^{\scr L} \Bigr)
		\biggr]
		\, .
\end{align}
Classically these symmetry generators are
observables, differing from the gauge-invariant ones~(\ref{Pi inv}) and~(\ref{Mij inv}) only off-shell.
Quantizing the gauge-fixed generators and requiring they remain observables 
produces the following operator ordering, according to Sec.~\ref{subsec: Quantum observables}:
\begin{align}
\hat{P}_i^\star ={}&
	\hat{P}_i^{\scr T}
	+
	\int\! d^{D-1}x \,
	\Bigl(
	\hat{K}_2^\dag \hat{A}_i^{\scr L} 
	+ \hat{A}_i^{\scr L} \hat{K}_2
	- \hat{K}_1^\dag \partial_i \hat{A}_0 
	- \partial_i \hat{A}_0 \hat{K}_1
	\Bigr)
	\, ,
\label{hatPi star}
\\
\hat{M}_{ij}^\star ={}&
	\hat{M}_{ij}^{\scr T}
	+
	\int\! d^{D-1}x \,
	\Bigl(
	2 \hat{K}_1^\dag x_{[i} \partial_{j]} \hat{A}_0
	+ 2 x_{[i} \partial_{j]} \hat{A}_0 \hat{K}_1
	- 2\hat{K}_2^\dag x_{[i}  \hat{A}_{j]}^{\scr L}
	- 2 x_{[i} \hat{A}_{j]}^{\scr L} \hat{K}_2
	\Bigr)
	\, ,
\label{hatMij star}
\end{align}
where the normal-ordered purely transverse parts were already given in~(\ref{PiT})
and~(\ref{MijT}). These quantum generators respect the~$E(D\!-\!1)$ algebra
at the operator level,
\begin{equation}
\bigl[ \hat{P}_i^\star , \hat{P}_j^\star \bigr] = 0 \, ,
\qquad
\bigl[ \hat{M}_{ij}^\star , \hat{P}_k^\star \bigr] = 2 i \hat{P}_{[i}^\star \delta_{j]k} \, ,
\qquad
\bigl[ \hat{M}_{ij}^\star , \hat{M}_{kl}^\star \bigr] =
	4 i  \delta_{i] [k} \hat{M}_{l] [j}^\star \, .
\end{equation}
In order for the translationally and rotationally invariant physical state to exist, it must be annihilated by both
symmetry generators~(\ref{hatPi star}) and~(\ref{hatMij star}), and by the non-Hermitian constraint~(\ref{K parametrized}).
This implies that the non-Hermitian constraint~$\mathcal{K}(\vec{k})$ must
commute with the gauge-fixed generators, modulo~$\hat{\mathcal{K}}$ itself.
That is already satisfied for translations,
\begin{equation}
\bigl[ \hat{\mathcal{K}}(\vec{k}) , \hat{P}_i^\star \bigr] 
	=
	k_i \, \hat{\mathcal{K}}(\vec{k})
	\, ,
\end{equation}
but only after requiring
\begin{equation}
\rho(\vec{k}) = \rho(k) \, ,
\qquad \qquad
\varphi(\vec{k}) = \varphi(k)
\, ,
\label{restrictions}
\end{equation}
is it satisfied for rotations,
\begin{equation}
\bigl[ \hat{\mathcal{K}}(\vec{k}) , \hat{M}_{ij}^\star \bigr] =
	i e^{i\theta(\vec{k})} 
	\Bigl( k_i \frac{\partial}{\partial k_j} \!-\! k_j \frac{\partial}{\partial k_i} \Bigr) 
		\Bigl( e^{-i\theta(\vec{k})}  \hat{\mathcal{K}}(\vec{k}) \Bigr) 
		\, .
\end{equation}
With the restrictions~(\ref{restrictions}) implemented 
in~(\ref{hatPi star}) and~(\ref{hatMij star}), the 
gauge-fixed generators take the form
\begin{align}
\hat{P}_i^\star ={}&
	\hat{P}_i^{\scr T}
	+
	\int\! d^{D-1}k \,
	k_i 
		\Bigl[ 
		\hat{\mathcal{K}}^\dag(\vec{k}) \hat{\mathcal{B}}(\vec{k})
		+
		\hat{\mathcal{B}}^\dag(\vec{k}) \hat{\mathcal{K}}(\vec{k})
		\Bigr]
		\, ,
\label{Pi star final}
\\
\hat{M}_{ij}^\star ={}&
	\hat{M}_{ij}^{\scr T}
	+
	\int\! d^{D-1}k \, (-i) \biggl[
	\Bigl( e^{i\theta(\vec{k})} \hat{\mathcal{B}}^\dag(\vec{k}) \Bigr)
	\Bigl( k_i \frac{\partial}{\partial k_j} \!-\! k_j \frac{\partial}{\partial k_i} \Bigr) 
	\Bigl( e^{-i\theta(\vec{k})} \hat{\mathcal{K}}(\vec{k}) \Bigr)
\nonumber \\
&	\hspace{2cm}
	+
	\Bigl( e^{i\theta(\vec{k})} \hat{\mathcal{K}}^\dag(\vec{k}) \Bigr)
	\Bigl( k_i \frac{\partial}{\partial k_j} \!-\! k_j \frac{\partial}{\partial k_i} \Bigr) 
	\Bigl( e^{-i\theta(\vec{k})} \hat{\mathcal{B}}(\vec{k}) \Bigr)
	\biggr]
	\, .
\label{Mij star final}
\end{align}
%


Next we turn to finding an eigenstate of gauge-fixed  generators~(\ref{Pi star final})
and~(\ref{Mij star final})
with vanishing eigenvalues.
This is now a condition on the scalar sector, 
since the transverse sector has already been fixed in 
Sec.~\ref{subsec: Physical symmetries}. 
Given that the subsidiary condition~(\ref{subsidiary K momentum})
acts on the scalar sector only,~$\hat{\mathcal{K}}(\vec{k}) \bigl| \Omega_0 \bigr\rangle\!=\!0$,
by simple inspection of generators~(\ref{Pi star final})
and~(\ref{Mij star final})
it is clear that the sought-for state vector has to be annihilated by~$\hat{\mathcal{B}}(\vec{k})$,
\begin{equation}
\hat{\mathcal{B}}(\vec{k}) \bigl| \Omega_0 \bigr\rangle = 0  \, ,
\label{B state condition}
\end{equation}
in addition to being annihilated by~$\hat{\mathcal{K}}(\vec{k})$. This guarantees that the
state is annihilated by the gauge-fixed symmetry generators,
\begin{equation}
\hat{P}_i^\star \bigl| \Omega \bigr\rangle = 0 \, ,
\qquad \qquad 
\hat{M}_{ij}^\star \bigl| \Omega \bigr\rangle = 0 \, .
\end{equation}
%


Having defined the homogeneous and isotropic state,
next we construct the scalar sector space of states. Since the operators acting on this 
vector space are~$\hat{\mathcal{K}}(\vec{k})$ and~$\hat{\mathcal{B}}(\vec{k})$,
the rest of the basis vectors are generated by acting with operators~$\hat{\mathcal{K}}^\dag(\vec{k})$
and~$\hat{\mathcal{B}}^\dag(\vec{k})$. However, this will not be a Fock space, since 
these operators are not the standard creation/annihilation operators due to their algebra.
There are several features to notice.

\bigskip

\noindent {\bf Indefinite metric (inner product) space.}
In the scalar sector the space of states is spanned 
by~$\hat{\mathcal{K}}^\dag(\vec{k})$ and~$\hat{\mathcal{B}}^\dag(\vec{k})$
acting on~$\bigl| \Omega_0 \bigr\rangle$.
In such a space there are states of vanishing and negative norm (in addition to 
positive norm). It is not difficult to construct examples. The two state vectors,
\begin{equation}
\bigl| \psi_1 \bigr\rangle = \int\! d^{D-1}k \, f(\vec{k}) \, \hat{\mathcal{K}}^\dag(\vec{k}) \bigl| \Omega_0 \bigr\rangle \, ,
\qquad
\bigl| \psi_2 \bigr\rangle = \int\! d^{D-1}k \, f(\vec{k}) \, \hat{\mathcal{B}}^\dag(\vec{k}) \bigl| \Omega_0 \bigr\rangle \, ,
\end{equation}
are orthogonal to each other,~$\bigl\langle \psi_1 \big| \psi_2 \bigr\rangle \!=\! 0$, but also both have a vanishing norm,
\begin{equation}
\bigl\langle \psi_1 \big| \psi_1 \bigr\rangle \!=\! 0 \, ,
\qquad \quad
\bigl\langle \psi_2 \big| \psi_2 \bigr\rangle \!=\! 0 \, .
\end{equation}
This is a consequence of commutation relations~(\ref{KB commutator}). 
It is also straightforward to demonstrate the
existence of negative norm states, e.g.
\begin{equation}
\bigl| \psi_3 \bigr\rangle 
	= \bigl| \psi_1 \bigr\rangle + \bigl| \psi_2 \bigr\rangle
\qquad \Longrightarrow \qquad
	\bigl\langle \psi_3 \big| \psi_3 \bigr\rangle =
	- 2 \int\! d^{D-1}k \, \bigl| f(\vec{k}) \bigr|^2 < 0 \, .
\end{equation}
Even though this might seem disconcerting at first, it is not really an issue,
as it does not affect the physical states defined by~(\ref{subsidiary K momentum}).

\bigskip

\noindent {\bf Physical subspace is positive-definite.}
The physical subspace of the entire space of states is defined by a subsidiary condition on 
the scalar sector space of states,~$\hat{\mathcal{K}}(\vec{k}) \bigl| \Omega_0^{\rm phys.} \bigr\rangle \!=\! 0$.
If the ``vacuum'' state of that subspace is defined by condition~(\ref{B state condition})
consistent with manifest homogeneity and isotropy, then it can be shown that 
the remaining members of the physical subspace take the form
\begin{equation}
\bigl| \Omega_0^{\rm phys.} \bigr\rangle = \bigl| \Omega_0 \bigr\rangle
	+ \sum_{n=1}^{\infty} \int\! d^{D-1}k_1 \dots d^{D-1}k_n \,
	f_n \bigl( \vec{k}_1 , \dots , \vec{k}_n \bigr) \,
	\hat{\mathcal{K}}^\dag(\vec{k}_1) \dots \hat{\mathcal{K}}^\dag(\vec{k}_n) 
	\bigl| \Omega_0 \bigr\rangle
	\, ,
\label{physical states}
\end{equation}
that all have a unit norm,
\begin{equation}
\bigl\langle \Omega_0^{\rm phys.} \big| \Omega_0^{\rm phys.} \bigr\rangle
	= \bigl\langle \Omega_0 \big| \Omega_0 \bigr\rangle = 1 \, .
\end{equation}
This form is dictated by the conditions~(\ref{subsidiary K momentum}) 
and~(\ref{B state condition}), and the algebra of operators~(\ref{KB commutator})
spanning the scalar sector space of states.
Physically there is no difference whatsoever which of the representatives in~(\ref{physical states})
we choose to represent the state. Therefore, the choice is delegated to a matter of convenience,
which is obviously the physical and homogeneous 
state.

\section{Two-point functions}
\label{sec: Two-point functions}

The two-point functions fully characterize Gaussian quantum states in free theories. 
Moreover, they are basic ingredients for nonequilibrium perturbative computations in 
field theory. In this section we first discuss state-independent properties of the 
two-point functions --- the equations of motion they satisfy and the various subsidiary conditions
they have to respect.
By the end of the section we express the photon two-point functions in terms of a few scalar mode functions
introduced in Sec.~\ref{sec: Field operator dynamics}. Thus, we reduce the future tasks
of computing photon two-point functions in FLRW spaces to computing several
scalar mode functions and the corresponding sum-over-modes.

\subsection{General properties}
\label{subsec: General properties}

The positive-frequency Wightman function for the photon is defined as an expectation value
of a product of two vector potential field operators,
\begin{equation}
i \bigl[ \tensor*[_\mu^{\scr - \!}]{\Delta}{_\nu^{\scr \! +}} \bigr] (x;x')
	=
	\bigl\langle \Omega \bigr| \hat{A}_\mu(x) \, \hat{A}_\nu(x') \bigl| \Omega \bigr\rangle
	\, ,
\label{photon Wightman}
\end{equation}
while the negative-frequency Wightman 
function,~$i \bigl[ \tensor*[_\mu^{\scr + \!}]{\Delta}{_\nu^{\scr \! -}} \bigr] (x;x')
	\!=\!
	\bigl\{ i \bigl[ \tensor*[_\mu^{\scr - \!}]{\Delta}{_\nu^{\scr \! +}} \bigr] (x;x') \bigr\}^*$,
is a complex conjugate that reverses the order of operators in the product in~(\ref{photon Wightman}).
These two can be used to define the Feynman propagator,
\begin{align}
i \bigl[ \tensor*[_\mu^{\scr + \!}]{\Delta}{_\nu^{\scr \! +}} \bigr] (x;x')
	={}&
	\bigl\langle \Omega \bigr| \mathcal{T} \Bigl( \hat{A}_\mu(x) \, \hat{A}_\nu(x') \Bigr) \bigl| \Omega \bigr\rangle
\nonumber \\
	={}&
	\theta(\eta \!-\! \eta' ) \, i \bigl[ \tensor*[_\mu^{\scr - \!}]{\Delta}{_\nu^{\scr \! +}} \bigr] (x;x')
	+
	\theta(\eta' \!-\! \eta ) \, i \bigl[ \tensor*[_\mu^{\scr + \!}]{\Delta}{_\nu^{\scr \! -}} \bigr] (x;x')
	\, ,
\label{Feynman def}
\end{align}
and its conjugate,~$i \bigl[ \tensor*[_\mu^{\scr - \!}]{\Delta}{_\nu^{\scr \! -}} \bigr] (x;x')
	\!=\!
	\bigl\{ i \bigl[ \tensor*[_\mu^{\scr + \!}]{\Delta}{_\nu^{\scr \! +}} \bigr] (x;x') \bigr\}^*$,
called the Dyson propagator.
The four two-point functions are completely determined by specifying the quantum state.
Nonetheless, there are general properties that they have to satisfy for any allowed
states. These properties are useful as checks of the consistency of two-point functions.
We derive and discuss them here.

First, the field operator equations of 
motion~(\ref{operator eom 1})--(\ref{operator eom 4}) can be written in a 
more familiar covariant form,
\begin{equation}
{\mathcal{D}_\mu }^\nu \hat{A}_\nu = 0 \, ,
\qquad
\mathcal{D}_{\mu\nu} = 
	g_{\mu\nu} \dalembertian
	- \nabla_\mu \nabla_\nu 
	+ \frac{1}{\xi} \bigl( \nabla_\mu \!+\! 2 \zeta n_\mu \bigr) \bigl(\nabla_\nu \!-\! 2 \zeta n_\nu \bigr)
	- R_{\mu\nu}
	\, .
\end{equation}
As a consequence  the Wightman function satisfies the same homogeneous equation of motion
on both external points,
\begin{equation}
{ \mathcal{D}_\mu }^\rho \, i \bigl[ \tensor*[_\rho^{\scr - \! }]{\Delta}{_\nu^{\scr \! +}} \bigr] (x;x') = 0 \, ,
\qquad \qquad
{ \mathcal{D}'_\nu }^\sigma \, i \bigl[ \tensor*[_\mu^{\scr - \! }]{\Delta}{_\sigma^{\scr \! +}} \bigr] (x;x') = 0 \, ,
\label{Wightman EOMs}
\end{equation}
and the canonical commutation relations~(\ref{ccr}) guarantee that the Feynman propagator 
satisfies inhomogeneous equations,
\begin{equation}
{ \mathcal{D}_\mu }^\rho \, i \bigl[ \tensor*[_\rho^{\scr + \! }]{\Delta}{_\nu^{\scr \! +}} \bigr] (x;x') 
	= g_{\mu\nu} \frac{i \delta^{D}(x\!-\!x')}{ \sqrt{-g} } \, ,
\qquad 
{ \mathcal{D}'_\nu }^\sigma \, i \bigl[ \tensor*[_\mu^{\scr + \! }]{\Delta}{_\sigma^{\scr \! +}} \bigr] (x;x') 
	= g_{\mu\nu} \frac{i \delta^{D}(x\!-\!x')}{ \sqrt{-g} } \, .
\label{Feynman EOMs}
\end{equation}
These are not the only state-independent equations that the photon two-point functions satisfy.
The quantization in Sec.~\ref{sec: Quantized photon in FLRW} required the two-point functions
of Hermitian constraints~(\ref{Hermitian two-points}) to vanish according to the correspondence principle.
By expressing the Hermitian constraints in terms of the derivatives of vector potential field operators,
\begin{equation}
\bigl( \nabla^\mu \!-\! 2 \zeta n^\mu \bigr) \hat{A}_\mu = \xi a^{2-D} \hat{\Pi}_0 
	= \xi a^{2-D} \hat{\Psi}_1
	\, ,
\qquad 
\bigl( 2g^{ij}   \delta^\mu_{[i} \partial_{0]} \partial_j \bigr) \hat{A}_{\mu}
	= a^{2-D} \partial_i \hat{\Pi}_i 
	= a^{2-D} \hat{\Psi}_2
	\, ,
\end{equation}
we can translate this quantization requirement into subsidiary conditions for the Wightman function,
\begin{subequations}
\begin{align}
\bigl( \nabla^\mu \!-\! 2 \zeta n^\mu \bigr) \bigl( \nabla'^\nu \!-\! 2 \zeta n'^\nu \bigr) 
	i \bigl[ \tensor*[_\mu^{\scr - \!}]{\Delta}{_\nu^{\scr \! +}} \bigr] (x;x') ={}& 0 \, ,
\label{double condition 1}
\\
\bigl( \nabla^\mu \!-\! 2 \zeta n^\mu \bigr)  \bigl( 2 g'^{kl} \delta^\nu_{[k} \partial'_{0]} \partial'_l \bigr) \,
	i \bigl[ \tensor*[_\mu^{\scr - \!}]{\Delta}{_{\nu}^{\scr \! +}} \bigr] (x;x') ={}& 0 \, ,
\label{double condition 2}
\\
\bigl( 2 g^{ij} \delta^\mu_{[i} \partial_{0]} \partial_j \bigr) \bigl( \nabla'^\nu \!-\! 2 \zeta n'^\nu \bigr) 
	 i \bigl[ \tensor*[_{\mu}^{\scr - \!}]{\Delta}{_\nu^{\scr \! +}} \bigr] (x;x') ={}& 0 \, ,
\label{double condition 3}
\\
\bigl( 2 g^{ij} \delta^\mu_{[i} \partial_{0]} \partial_j \bigr)
	\bigl( 2 g'^{kl} \delta^\nu_{[k} \partial'_{0]} \partial'_l \bigr) \,
	i \bigl[ \tensor*[_{\mu}^{\scr - \!}]{\Delta}{_{\nu}^{\scr \! +}} \bigr] (x;x') ={}& 0 \, ,
\label{double condition 4}
\end{align}
\label{all double conditions W}%
\end{subequations}
and for the Feynman propagator,
\begin{subequations}
\begin{align}
\bigl( \nabla^\mu \!-\! 2 \zeta n^\mu \bigr) \bigl( \nabla'^\nu \!-\! 2 \zeta n'^\nu \bigr) 
	i \bigl[ \tensor*[_\mu^{\scr + \!}]{\Delta}{_\nu^{\scr \! +}} \bigr] (x;x') 
	={}&
	- \xi \frac{i \delta^D(x\!-\!x')}{ \sqrt{-g} }
	\, ,
\label{double condition F 1}
\\
\bigl( \nabla^\mu \!-\! 2 \zeta n^\mu \bigr)  \bigl( 2 g'^{kl} \delta^\nu_{[k} \partial'_{0]} \partial'_l \bigr) \,
	i \bigl[ \tensor*[_\mu^{\scr + \!}]{\Delta}{_{\nu}^{\scr \! +}} \bigr] (x;x') 
	={}&
	0 
	\, ,
\\
\bigl( 2 g^{ij} \delta^\mu_{[i} \partial_{0]} \partial_j \bigr) 
	\bigl( \nabla'^\nu \!-\! 2 \zeta n'^\nu \bigr) 
	 i \bigl[ \tensor*[_{\mu}^{\scr + \!}]{\Delta}{_\nu^{\scr \! +}} \bigr] (x;x') 
	 ={}&
	 0 
	 \, ,
\\
\bigl( 2 g^{ij} \delta^\mu_{[i} \partial_{0]} \partial_j \bigr)
	\bigl( 2 g'^{kl} \delta^\nu_{[k} \partial'_{0]} \partial'_l \bigr) \,
	i \bigl[ \tensor*[_{\mu}^{\scr + \!}]{\Delta}{_{\nu}^{\scr \! +}} \bigr] (x;x') 
	={}&
	\partial_i \partial'_i \, \frac{ i\delta^D(x\!-\!x')}{\sqrt{-g}}
	\, .
\end{align}
\label{all double conditions F}%
\end{subequations}
These are useful as consistency checks of two-point functions.
Failure to satisfy them signals inconsistencies of photon two-point functions.

\medskip

The two-derivative subsidiary conditions~(\ref{all double conditions W})
and~(\ref{all double conditions F})
are independent of the choice of state, and in particular of the choice
of the pure gauge sector. There is a different way of expressing~(\ref{all double conditions W}) 
and~(\ref{all double conditions F}),
in terms of single-derivative subsidiary conditions. These derive from considering correlators 
between Hermitian 
constraints and vector field operators,
\begin{align}
\bigl( \nabla^\mu \!-\! 2 \zeta n^\mu \bigr)
	i \bigl[ \tensor*[_\mu^{\scr - \!}]{\Delta}{_\nu^{\scr \!+} } \bigr] (x;x')
	={}&
	\xi a^{2-D} \bigl\langle \Omega \bigr| \hat{\Psi}_1(x) \hat{A}_\nu(x') \bigl| \Omega \bigr\rangle 
	\, ,
\\
\bigl( 2 g^{ij} \delta^\mu_{[i} \partial_{0]} \partial_j \bigr) 
	i \bigl[ \tensor*[_\mu^{\scr - \!}]{\Delta}{_\nu^{\scr \!+} } \bigr] (x;x')
	={}&
	a^{2-D} \bigl\langle \Omega \bigr| \hat{\Psi}_2(x) \hat{A}_\nu(x') \bigl| \Omega \bigr\rangle 
	\, .
\end{align}
Given the decomposition~(\ref{Hermitian decomposition}) of Hermitian constraints into non-Hermitian ones, 
and the subsidiary condition in the state~(\ref{K position condition}),
the right-hand sides above reduce to,
\begin{align}
\bigl( \nabla^\mu \!-\! 2 \zeta n^\mu \bigr)
	i \bigl[ \tensor*[_\mu^{\scr - \!}]{\Delta}{_\nu^{\scr \!+} } \bigr] (x;x')
	={}&
	\xi a^{2-D} \bigl\langle \Omega \bigr| \bigl[ \hat{K}_1(x) , \hat{A}_\nu(x') \bigr] \bigl| \Omega \bigr\rangle 
\label{subs 1}
	\, ,
\\
\bigl( 2 g^{ij} \delta^\mu_{[i} \partial_{0]} \partial_j \bigr) 
	i \bigl[ \tensor*[_\mu^{\scr - \!}]{\Delta}{_\nu^{\scr \!+} } \bigr] (x;x')
	={}&
	a^{2-D} \bigl\langle \Omega \bigr| \bigl[ \hat{K}_2(x) , \hat{A}_\nu(x') \bigr] \bigl| \Omega \bigr\rangle 
	\, .
\label{subs 2}
\end{align}
Evaluating the position space commutators 
on the right-hand side above is simpler if we first compute the momentum space commutators,
\begin{align}
\bigl[ \hat{\mathcal{K}}(\vec{k}) , \hat{\mathcal{A}}_0(\eta, \vec{k}^{\,\prime}) \bigr]
	={}&
	- e^{i \theta(\vec{k})}
	\mathscr{U}_\lambda^*(\eta,k) \,
	\delta^{D-1} ( \vec{k} \!+\! \vec{k}^{\,\prime} )
	\, ,
\label{KA0 commutator}
\\
\bigl[ \hat{\mathcal{K}}(\vec{k}) , \hat{\mathcal{A}}_{\scr L}(\eta, \vec{k}^{\,\prime}) \bigr]
	={}&
	- i e^{i \theta(\vec{k})}
	\mathscr{U}_{\lambda+1}^*(\eta,k)
	\delta^{D-1} ( \vec{k} \!+\! \vec{k}^{\,\prime} )
	\, ,
\label{KAL commutator}
\end{align}
where we defined the two scalar mode functions,
\begin{align}
\mathscr{U}_\lambda(\eta,k) ={}&
	e^{i \varphi(k)} {\rm ch}[\rho(k)] \, \mathcal{U}_\lambda(\eta,k)
	- 
	e^{-i \varphi(k)} {\rm sh}[\rho(k)] \, \mathcal{U}_\lambda^*(\eta,k)
	\, ,
\label{UL}
\\
\mathscr{U}_{\lambda+1}(\eta,k) ={}&
	e^{i \varphi(k)} {\rm ch}[\rho(k)] \, \mathcal{U}_{\lambda+1}(\eta,k)
	+
	e^{-i \varphi(k)} {\rm sh}[\rho(k)] \, \mathcal{U}_{\lambda+1}^*(\eta,k)
	\, ,
\label{UL+1}
\end{align}
that also satisfy the  recurrence relations~(\ref{recurrences}),
\begin{equation}
\!\!\!\!
\biggl[ \partial_0 + \! \Bigl( \lambda \!+\! \frac{1}{2} \Bigr)(1\!-\!\epsilon) \mathcal{H} \biggr]
	\mathscr{U}_\lambda
	\!=\! - i k \, \mathscr{U}_{\lambda+1} \, ,
\quad 
\biggl[ \partial_0 - \! \Bigl( \lambda \!+\! \frac{1}{2} \Bigr)(1\!-\!\epsilon) \mathcal{H} \biggr]
	\mathscr{U}_{\lambda+1}
	\!=\! - i k \, \mathscr{U}_{\lambda} \, .
\end{equation}
Using these recurrence relations, the commutators~(\ref{KA0 commutator}) and~(\ref{KAL commutator}),
Fourier transforms~(\ref{Fouriers}) and~(\ref{Kfours}), and the operators~(\ref{K1K2}),
it is straightforward to show that the right-hand sides of~(\ref{subs 1})
and~(\ref{subs 2}) are
\begin{align}
\bigl( \nabla^\mu \!-\! 2 \zeta n^\mu \bigr)
	i \bigl[ \tensor*[_\mu^{\scr - \!}]{\Delta}{_\nu^{\scr \!+} } \bigr] (x;x')
	={}&
	- \xi \partial_\nu' \biggl\{
	\Bigl( \frac{a'}{a} \Bigr)^{\! \zeta} 
	i \bigl[ \tensor*[^{\scr \!-\!}]{\Delta}{^{\scr \!+\!} } \bigr]_{\lambda+1} (x;x')
	\biggr\}
	\, ,
\label{single condition 1}
\\
\bigl( 2 g^{ij} \delta^\mu_{[i} \partial_{0]} \partial_j \bigr) 
	i \bigl[ \tensor*[_\mu^{\scr - \!}]{\Delta}{_\nu^{\scr \!+} } \bigr] (x;x')
	={}&
	- \bigl( \partial_0 \!+\! 2 \zeta \mathcal{H} \bigr) \partial'_\nu
		\biggl\{
		\Bigl( \frac{a'}{a} \Bigr)^{\! \zeta} i \bigl[ \tensor*[^{\scr \!-\!} ]{\Delta}{ ^{\scr \!+\!} } \bigr]_{\lambda+1} (x;x')
		\biggr\}
\label{single condition 2}
	\, ,
\end{align}
where we recognized the scalar two-point function~(\ref{scalar two-point}),
\begin{align}
\MoveEqLeft[4]
i \bigl[ \tensor*[^{\scr \!-\!}]{\Delta}{^{\scr \!+\!} } \bigr]_{\lambda+1} (x;x')
	=
	(aa')^{-\frac{D-2}{2}} 
	\! \int\! \frac{ d^{D-1}k }{ (2\pi)^{D-1} } \,
	e^{i \vec{k} \cdot (\vec{x} - \vec{x}^{\,\prime} )}
	\mathscr{U}_{\lambda+1}(\eta,k)
	\mathscr{U}_{\lambda+1}^*(\eta',k)
	\, .
\end{align}
It satisfies the scalar two-point function equation of motion~(\ref{scalar two-point function eom}),
which in turn guarantees that conditions~(\ref{all double conditions W}) are satisfied.
Analogous results for the Feynman propagator are derived in the same manner,
\begin{align}
\bigl( \nabla^\mu \!-\! 2 \zeta n^\mu \bigr)
	i \bigl[ \tensor*[_\mu^{\scr + \!}]{\Delta}{_\nu^{\scr \!+} } \bigr] (x;x')
	={}&
	- \xi \partial_\nu' \biggl\{
	\Bigl( \frac{a'}{a} \Bigr)^{\! \zeta} 
	i \bigl[ \tensor*[^{\scr \!+\!}]{\Delta}{^{\scr \!+\!} } \bigr]_{\lambda+1} (x;x')
	\biggr\}
	\, ,
\label{single condition 1 F}
\\
\bigl( 2 g^{ij} \delta^\mu_{[i} \partial_{0]} \partial_j \bigr) 
	i \bigl[ \tensor*[_\mu^{\scr + \!}]{\Delta}{_\nu^{\scr \!+} } \bigr] (x;x')
	={}&
	- \bigl( \partial_0 \!+\! 2 \zeta \mathcal{H} \bigr) \partial'_\nu
		\biggl\{
		\Bigl( \frac{a'}{a} \Bigr)^{\! \zeta} i \bigl[ \tensor*[^{\scr \!+\!} ]{\Delta}{ ^{\scr \!+\!} } \bigr]_{\lambda+1} (x;x')
		\biggr\}
		+
		\delta_\nu^0 \frac{ i \delta^D(x\!-\!x')}{ a^{D-2}}
\label{single condition 2 F}
	\, ,
\end{align}
and the scalar Feynman propagator satisfies~(\ref{scalar Feynman eom}),
which guarantees conditions~(\ref{all double conditions F}) hold.
Expressions~(\ref{single condition 1}) and~(\ref{single condition 2}) for the Wightman function,
and~(\ref{single condition 1 F}) and~(\ref{single condition 2 F}) for the Feynman propagator
are the promised single-derivative subsidiary conditions.

This is the point where the formalism outlined in this work makes connection with the 
Becchi-Rouet-Stora-Tyutin (BRST) quantization and Faddeev-Popov (FP) ghosts.
On the right-hand sides of the expressions for the single derivative conditions above 
we can recognize the FP ghost two-point function
\begin{equation}
\bigl\langle \overline{c}(x) c(x') \bigr\rangle
	= \Bigl( \frac{a'}{a} \Bigr)^{\!\zeta} i \bigl[ \tensor*[^{\scr \!-\!}]{\Delta}{^{\scr \!+\!} } \bigr]_{\lambda+1} (x;x') \, ,
\label{FP propagator}
\end{equation}
that satisfies the equation of motion
\begin{equation}
\bigl( \nabla'^\mu \!-\! 2\zeta n'^\mu \bigr) \nabla'_\mu \bigl\langle \overline{c}(x) c(x') \bigr\rangle
	= 0 \, ,
\label{ghost equation}
\end{equation}
such that the single derivative subsidiary condition~(\ref{single condition 1}) takes the form
of the Ward-Takahashi identity,
\begin{equation}
\bigl( \nabla^\mu \!-\! 2 \zeta n^\mu \bigr)
	i \bigl[ \tensor*[_\mu^{\scr - \!}]{\Delta}{_\nu^{\scr \!+} } \bigr] (x;x')
	=
	- \xi \partial_\nu' \bigl\langle \overline{c}(x) c(x') \bigr\rangle \, .
\label{WT identity}
\end{equation}
This is precisely the condition that descends from the BRST quantization.
In addition to the gauge-fixing action functional~(\ref{gauge-fixing term}), 
there one introduces the accompanying
FP ghost action for Grassmanian fields~$c$ and~$\overline{c}$,
\begin{equation}
S_{\rm gh}[\overline{c},c] = \int\! d^{D\!}x \, \sqrt{-g} \, 
	\bigl( \nabla^\mu \overline{c} + 2\zeta n^\mu \overline{c} \bigr) 
	\bigl( \nabla_\mu c \bigr)
	\, .
\label{ghost action}
\end{equation}
such that the total action is invariant under global BRST transformations,
\begin{equation}
A_\mu \longrightarrow A_\mu + \theta \xi \partial_\mu c \, ,
\qquad\qquad
\overline{c} \longrightarrow \overline{c} - \theta \bigl( \nabla^\mu \!-\! 2 \zeta n^\mu \bigr) A_\mu \, ,
\qquad\qquad
c \longrightarrow c \, ,
\end{equation}
parametrized by an infinitesimal Grassmanian parameter~$\theta$. This implies a conserved BRST charge
generating the transformation. The physical states in BRST quantization 
are required to be invariant under the action of this charge, which 
yields the Ward-Takahashi identity~(\ref{WT identity}) as a consequence.
It is Eq.~(\ref{ghost equation}) that guarantees the single-derivative 
subsidiary conditions ~(\ref{single condition 1}) and~(\ref{single condition 2}) are
consistent with the vanishing of the double-derivative 
subsidiary conditions~(\ref{double condition 1})--(\ref{double condition 4}).
Thus the correlators of Hermitian constraints vanish, and the
correspondence principle is respected.

\subsection{Mode sum representation}
\label{subsec: Mode sum representation}

The components of the photon two-point functions can be expressed as integrals over modes
of products of mode functions. This is accomplished by using conditions imposed on the state both 
in the physical sector~(\ref{transverse vacuum})
and in the gauge sector in~(\ref{subsidiary K momentum}) and~(\ref{B state condition}), 
and the momentum space representation of the field operators~(\ref{Fouriers}),
\begin{align}
&
i \bigl[ \tensor*[_0^{\scr - \!}]{\Delta}{_0^{\scr \! +}} \bigr] (x;x')
	=
\label{mode sum 00}
\\
&	\hspace{0.9cm}
	(aa')^{ - \frac{D-2-2\zeta}{2}} \!\! 
	\int\! \frac{ d^{D-1}k }{ (2\pi)^{D-1} } \, e^{i \vec{k} \cdot (\vec{x} - \vec{x}^{\,\prime} )}
	\Bigl[
	- \mathscr{U}_{\lambda}(\eta,k) \mathscr{V}_0^*(\eta',k)
	-
	\mathscr{V}_0(\eta,k) \mathscr{U}_{\lambda}^*(\eta',k)
	\Bigr]
	\, ,
\nonumber \\
&
i \bigl[ \tensor*[_0^{\scr - \!}]{\Delta}{_i^{\scr \! +}} \bigr] (x;x') =
\label{mode sum 0i}
\\
&	\hspace{0.9cm}
	(aa')^{ - \frac{D-2-2\zeta}{2}} \!\! 
	\int\! \frac{ d^{D-1}k }{ (2\pi)^{D-1} } \, e^{i \vec{k} \cdot (\vec{x} - \vec{x}^{\,\prime} )}
	\, \frac{k_i}{k}
	\Bigl[
	\mathscr{U}_\lambda(\eta,k) \mathscr{V}_{\scr L}^*(\eta',k)
	+ 
	\mathscr{V}_0(\eta,k) \mathscr{U}_{\lambda+1}^*(\eta',k)
	\Bigr]
	\, ,
\nonumber \\
&
i \bigl[ \tensor*[_{i\,}^{\scr - \!}]{\Delta}{_j^{\scr \! +}} \bigr] (x;x')
	= \!
	(aa')^{-\frac{D-4}{2}} \!\!
	\int\! \frac{ d^{D-1}k }{ (2\pi)^{D-1} } \, e^{i \vec{k} \cdot (\vec{x} - \vec{x}^{\,\prime} )}
	\Bigl( \delta_{ij} \!-\! \frac{k_i k_j}{k^2} \Bigr) \, \mathscr{U}_{\nu_{\scr T}}(\eta,k) \, \mathscr{U}_{\nu_{\scr T}}^*(\eta',k)
\label{mode sum ij}
\\
&
	-
	(aa')^{ - \frac{D-2-2\zeta}{2}} \!\! 
	\int\! \frac{ d^{D-1}k }{ (2\pi)^{D-1} } \, e^{i \vec{k} \cdot (\vec{x} - \vec{x}^{\,\prime} )}
	\, \frac{k_i k_j}{k^2}
	\Bigl[
	\mathscr{U}_{\lambda+1}(\eta,k) \mathscr{V}_{\scr L}^*(\eta',k)
	+
	\mathscr{V}_{\scr L}(\eta,k) \mathscr{U}_{\lambda+1}^*(\eta',k)
	\Bigr]
	\, ,
\nonumber
\end{align}
where we introduced the following shorthand notation for scalar mode functions,
\begin{align}
\mathscr{U}_{\nu_{\scr T}}(\eta,k) ={}& 
	e^{i \varphi_{\scr T}(k)} {\rm ch}[\rho_{\scr T}(k)] \, \mathcal{U}_{\nu_{\scr T}}(\eta,k)
	-
	e^{-i \varphi_{\scr T}(k)} {\rm sh}[\rho_{\scr T}(k)] \, \mathcal{U}_{\nu_{\scr T}}^*(\eta,k)
	\, ,
\\
\mathscr{V}_0(\eta,k) ={}&
	e^{i \varphi(k) } {\rm ch}[\rho(k)] v_0(\eta,k)
		- e^{- i \varphi(k)} {\rm sh}[\rho(k)] v_0^*(\eta,k)
		\, ,
\\
\mathscr{V}_{\scr L}(\eta,k) ={}&
	e^{i \varphi(k)} {\rm ch}[\rho(k)] \, v_{\scr L}(\eta,k) 
		+ e^{-i\varphi(k)} {\rm sh}[\rho(k)] \, v_{\scr L}^*(\eta,k) 
		\, ,
\end{align}
in addition to the ones already defined in~(\ref{UL}) and~(\ref{UL+1}).
The Wightman function constructed this way is guaranteed to satisfy both the equations of 
motion~(\ref{Wightman EOMs}) {\it and} the appropriate subsidiary conditions~(\ref{all double conditions W}).
The Feynman propagator follows from the Wightman function simply from the definition~(\ref{Feynman def})
and satisfies the equations of motion~(\ref{Feynman EOMs}) and subsidiary conditions~(\ref{all double conditions F}).
This guarantees that the perturbation theory
based on these two-point functions will yield correct results.

\section{Simple observables}
\label{sec: Simple observables}

The two-point functions computed according to the preceding section can be used
to compute quantum loop corrections to various observables in spatially flat
cosmological spaces. However, in addition to working out the two-point functions,
we also have to address the question of ordering products of field operators 
comprising the observables. This is very much related to the fact that observables have
to be independent of the gauge-fixing parameter~$\xi$.
To elucidate this point, in this section we consider two
simple observables: the tree-level field strength correlator and the one-loop 
energy-momentum tensor.

\subsection{Field strength correlators}
\label{subsec: Field strength correlators}

The simplest observable one can think of is the tree-level off-coincident field strength correlator,
\begin{equation}
\bigl\langle \Omega \bigr| \hat{F}_{\mu\nu}(x) \hat{F}_{\rho\sigma}(x') \bigl| \Omega \bigr\rangle
	= 4 \bigl( \delta^\alpha_{[\mu} \partial_{\nu]} \bigr) 
		\bigl( \delta^\beta_{[\rho} \partial'_{\sigma]} \bigr)
		i \bigl[ \tensor*[_\alpha^{\scr - \!}]{\Delta}{_\beta^{\scr \! +}} \bigr](x;x') \, ,
\label{F correlator def}
\end{equation}
expressed in terms of derivatives acting on the vector potential two-point function.
The field strength tensor is an observable, as it fully commutes with the Hermitian
constraints, and thus also with the non-Hermitian constraint. This is obvious if we write it in
terms of canonical variables of the transverse and scalar sectors,
\begin{equation}
\hat{F}_{0i} = a^{4-D} \hat{\Pi}_i^{\scr T} + a^{4-D} \hat{\Pi}_i^{\scr L} \, ,
\qquad \qquad
\hat{F_{ij}} = 2 \partial_{[i} \hat{A}_{j]}^{\scr T} \, .
\end{equation}
This also makes it clear that the correlator~(\ref{F correlator def}) receives contributions from 
the transverse sector only, and that any gauge-dependence drops out,
\begin{align}
\bigl\langle \Omega \bigr| \hat{F}_{0i}(x) \hat{F}_{0j}(x') \bigl| \Omega \bigr\rangle 
	={}& \partial_0 \partial'_0 \bigl\langle \Omega \bigr| \hat{A}_i^{\scr T}(x) \hat{A}_j^{\scr T}(x') 
		\bigl| \Omega  \bigr\rangle \, ,
\\
\bigl\langle \Omega \bigr| \hat{F}_{0i}(x) \hat{F}_{kl}(x') \bigl| \Omega \bigr\rangle 
	={}&
	2 \partial_0 \bigl( \delta_{[l}^n \partial'_{k]} \bigr) 
		\bigl\langle \Omega \bigr| \hat{A}_i^{\scr T}(x) \hat{A}_{n}^{\scr T}(x') 
		\bigl| \Omega  \bigr\rangle 
	\, ,
\\
\bigl\langle \Omega \bigr| \hat{F}_{ij}(x) \hat{F}_{kl}(x') \bigl| \Omega \bigr\rangle 
	={}&
	4 \bigl( \delta_{[j}^m \partial_{i]} \bigr) \bigl( \delta_{[l}^n \partial'_{k]} \bigr) 
	 \bigl\langle \Omega \bigr| \hat{A}_m^{\scr T}(x) \hat{A}_n^{\scr T}(x') 
		\bigl| \Omega  \bigr\rangle 
	\, .
\end{align}
The specific forms that these correlators take depend on the particular FLRW spacetime,
and on the state of the transverse modes that is chosen, i.e. on the free coefficients
chosen in~(\ref{cT}). In the four-dimensional limit the transverse photons are conformally
coupled, and there exists a conformal vacuum state, that has to reproduce the flat space 
correlators,
\begin{equation}
\bigl\langle \Omega \bigr| \hat{F}_{\mu\nu}(x) \hat{F}_{\rho\sigma}(x') \bigl| \Omega \bigr\rangle
	\xrightarrow{D\to4}
	\frac{2}{\pi^2 \bigl(\Delta x^2 \bigr)^{\!2} }
	\biggl[
	\eta_{\mu[\rho} \eta_{\sigma]\nu} 
	- 4 \eta_{\alpha [\mu} \eta_{\nu] [\sigma } \eta_{\rho] \beta}
		\frac{ \Delta x^\alpha \Delta x^\beta }{ \Delta x^2}
	\biggr]
	\, .
\end{equation}
This is a simple check that any photon two-point function of a physically conformal vacuum state in FLRW
space must satisfy.

\subsection{Energy-momentum tensor}
\label{subsec: Energy-momentum tensor}

The energy-momentum tensor is perhaps the simplest one-loop observable,
composed of a single photon two-point function. There is an ambiguity,
even at the classical level, in how we define 
even the observable, which consequently appears in the quantized theory as well. This ambiguity,
however, vanishes on-shell both in the classical and in the quantum cases. Most of this
subsection is devoted to the discussion of how to properly define the quantum energy-momentum tensor.

\subsubsection{Classical energy-momentum tensor}
\label{subsubsec: Classical energy-momentum tensor}

Two sensible definitions for the energy-momentum tensor of the photon field are possible.
It can be defined either as a variation of the gauge-invariant action~(\ref{invariant action}),
\begin{equation}
T_{\mu\nu} = \frac{-2}{\sqrt{-g}} \frac{\delta S}{\delta g^{\mu\nu}}
	= \Bigl( \delta_\mu^\rho \delta_\nu^\sigma - \frac{1}{4} g_{\mu\nu} g^{\rho\sigma} \Bigr) 
		g^{\alpha\beta} F_{\rho\alpha} F_{\sigma\beta}
		\, ,
\label{T def}
\end{equation}
or as a variation of the gauge-fixed action~(\ref{fixed action}),
\begin{equation}
T_{\mu\nu}^\star = \frac{-2}{ \sqrt{-g} } \frac{\delta S_\star}{\delta g^{\mu\nu}}
	= T_{\mu\nu} + T_{\mu\nu}^{\rm gf}
	\, ,
\end{equation}
which, in addition to the gauge-invariant part, contains an extra contribution from the gauge-fixing term~(\ref{gauge-fixing term}),
\begin{align}
T_{\mu\nu}^{\rm gf} 
	={}& \frac{-2}{\sqrt{-g}} \frac{\delta S_{\rm gf} }{ \delta g^{\mu\nu} }
	=
	- \frac{ 2 }{\xi} 
		\Bigl( A_{(\mu} \nabla_{\nu)} + 2 \zeta n_{(\mu} A_{\nu)} \Bigr)
		\Bigl( \nabla^\rho A_\rho - 2 \zeta n^\rho A_\rho \Bigr)
\label{Tgf def}
\\
&	\hspace{1cm}
	+
	\frac{ g_{\mu\nu} }{\xi} 
		\biggl[
		\Bigl( A_\rho \nabla^\rho + 2 \zeta n^\rho A_\rho \Bigr)
		\Bigl( \nabla^\sigma A_\sigma - 2 \zeta n^\sigma A_\sigma \Bigr)
	+
	\frac{1}{2} \Bigl( \nabla^\rho A_\rho
		- 2 \zeta n^\rho A_\rho \Bigr)^{\!2} \,
	\biggr]
	\, .
\nonumber 
\end{align}
Both definitions give energy-momentum tensors conserved on-shell, but for the first 
definition~(\ref{T def}) we have to use the constraint equations~(\ref{const eq}), while the conservation
for the second definition~(\ref{Tgf def}) relies on the gauge-fixed dynamical 
equations~(\ref{fixed EOM1})--(\ref{fixed EOM4}) only.
At the classical level the two definitions give the same answer on-shell.
This is best seen by expressing the gauge-fixing contribution in terms of canonical variables,
\begin{subequations}
\begin{align}
T^{\rm gf}_{00}	
	={}&
		- a^{2-D} \biggl[
			A_0 \Psi_2
			+
			A_k \partial_k \Psi_1
			+
			\frac{\xi}{2} a^{4-D} \Psi_1^2
		\biggr]
		\, ,
\\
T^{\rm gf}_{0i}	
	={}& 
	- a^{2-D} \Bigl[
		A_0 \partial_i \Psi_1
		+
		A_i \Psi_2
	\Bigr]
	\, ,
\\
T^{\rm gf}_{ij}	
	={}& 
	- a^{2-D}\biggl[
		2 A_{(i} \partial_{j)} \Psi_1
		+
		\delta_{ij}
		\biggl(
		 A_0 \Psi_2
		-
		A_k \partial_k \Psi_1
		-
		\frac{\xi}{2} a^{4-D} \Psi_1^{2}
		\biggr)
	\biggr]
	\, ,
\end{align}
\label{Tgf}%
\end{subequations}
and noting that every term contains at least one of the first-class constraints~(\ref{on-shell eq5}).
In fact it is only the transverse modes that contribute to the energy-momentum tensor
on-shell. This is clear since the only contributing part is the gauge-invariant one~(\ref{T def})
that is composed out of field strength tensors only, which contain only transverse fields
and constraints, as discussed in Sec.~\ref{subsec: Field strength correlators}.
The same properties of the energy-momentum tensor are maintained in the quantized 
theory if attention is paid to operator ordering.

%
%
%
%
%
%

\subsubsection{Quantum energy-momentum tensor}
\label{subsubsec: quantum energy-momentum tensor}

When defining operators associated with quantum observables attention needs to be paid to
the ordering of products of field operators. Usually Weyl ordering of field operators 
is employed.
However, this is not fully satisfactory in gauge theories as, in general, it does not respect
the correspondence principle. For the energy-momentum tensor this is the question of the
contribution of the gauge-fixing part~(\ref{Tgf def}). In the classical theory this contribution
vanishes, and it is sensible to demand the same property in the quantized theory.
This is accomplished by correct operator ordering. Here we discuss the quantization of the
two parts of the definition~(\ref{T def}) and~(\ref{Tgf def}) separately.

\bigskip

\noindent {\bf Gauge-invariant part.} 
For the gauge-invariant part~(\ref{T def}) defining an operator is straightforward,
since when expressed in terms of the canonical fields all the terms are composed 
either solely of transverse fields or solely of constraints. Therefore, we may 
define the operator 
to be Weyl-ordered, and the expectation value essentially reduces to the coincident 
limit of the field strength correlator,
\begin{equation}
\bigl\langle \Omega \bigr| \hat{T}_{\mu\nu}(x) \bigl| \Omega \bigr\rangle
	=
	\Bigl( \delta_\mu^\rho \delta_\nu^\sigma - \frac{1}{4} g_{\mu\nu} g^{\rho\sigma} \Bigr) 
		g^{\alpha\beta} 
		\times \frac{1}{2} 
		\bigl\langle \Omega \bigr| \bigl\{ \hat{F}_{\rho\alpha}(x) ,
			\hat{F}_{\sigma\beta}(x) \bigr\} \bigl| \Omega \bigr\rangle
		\, .
\label{T expectation}
\end{equation}
If the off-coincident field strength correlator~(\ref{F correlator def}) is computed in~$D$ dimensions
we can take a dimensionally regulated coincidence limit required above.  The precise value of~(\ref{T expectation})
depends on the transverse photon state and the particular FLRW background.

\bigskip

\noindent {\bf Gauge-fixing part.} 
The gauge-fixing contribution~(\ref{Tgf def}) to the energy-momentum 
tensor contains constraints, as evident from~(\ref{Tgf}).
Therefore, the operator associated with the gauge-fixing part of the energy-momentum 
tensor has to be ordered properly, as explained in Sec.~\ref{subsec: Quantum observables}. 
The Hermitian constraints have to be split into parts containing the non-Hermitian 
subsidiary constraint~$\hat{K}$ and parts containing its conjugate~$\hat{K}^\dag$, 
and the former has to be put to the right of the product and the latter to the left.
This is accomplished by making use of decompositions in~(\ref{Hermitian decomposition}),
\begin{subequations}
\begin{align}
\hat{T}^{\rm gf}_{00}	
	\equiv \Bigl[ \hat{T}^{\rm gf}_{00} \Bigr]_g
	={}&
	- a^{2-D} \biggl[
	\Bigl(
		\hat{K}_2^\dag \hat{A}_{0}
		+ \partial_k \hat{K}_1^\dag \hat{A}_k
		\Bigr)
	+ \Bigl(
		\hat{A}_{0} \hat{K}_2
		+ \hat{A}_k \partial_k \hat{K}_1
		\Bigr)
	+ \frac{\xi}{2} a^{4-D} \hat{\Psi}_1^2
	\biggr]
	\, ,
\label{T00 operator}
\\
\hat{T}^{\rm gf}_{0i}	
	\equiv \Bigl[ \hat{T}^{\rm gf}_{0i} \Bigr]_g
	={}&
	- 
	a^{2-D}
	\biggl[
	\Bigl(
		\partial_{i} \hat{K}_1^\dag \hat{A}_{0}
		+ \hat{K}_2^\dag \hat{A}_{i}
	\Bigr)
	+ \Bigl(
		\hat{A}_{0} \partial_{i} \hat{K}_1
		+ \hat{A}_{i} \hat{K}_2
	\Bigr)
	\biggr]
	\, ,
\label{T0i operator}
\\
\hat{T}^{\rm gf}_{ij}	
	\equiv \Bigl[ \hat{T}^{\rm gf}_{ij} \Bigr]_g
	={}& 
	-
	a^{2-D}
	\biggl[
	2 \partial_{(i} \hat{K}_1^\dag \hat{A}_{j)} 
	+ \delta_{ij} \Bigl(
		\hat{K}_2^\dag \hat{A}_0 
		- \partial_k \hat{K}_1^\dag \hat{A}_k \Bigr) 
\nonumber \\
&	\hspace{2cm}
	+
	2 \hat{A}_{(i} \partial_{j)} \hat{K}_1
	+ \delta_{ij} \Bigl(
		\hat{A}_0 \hat{K}_2 
		- \hat{A}_k \partial_k \hat{K}_1 \Bigr)
	- \frac{\xi}{2} \delta_{ij} a^{4-D} \hat{\Psi}_1^2
	\biggr]
	\, .
\label{Tij operator}
\end{align}
\label{Tmunu operator}%
\end{subequations}
Note that for terms composed solely of constraints this ordering is immaterial
since the non-Hermitian subsidiary constraint and its conjugate commute.
Defined this way it is manifest that the expectation value always vanishes,
\begin{equation}
\bigl\langle \Omega \bigr| \hat{T}_{\mu\nu}^{\rm gf}(x) \bigl| \Omega \bigr\rangle = 0 \, ,
\label{Tgf vanishing}
\end{equation}
as it should, without any additional arguments.
%
%

Even though the operator ordering in~(\ref{Tmunu operator}) leads to correct and consistent results,
it is rather unwieldy to use it in practice. In general Weyl-ordered products are far more convenient 
to use.
We can indeed commute the operators in products of~(\ref{Tmunu operator}) 
to the Weyl-ordered form, but this leaves nonvanishing commutators accounting for the 
difference between two ordering prescriptions,
\begin{subequations}
\begin{align}
\Bigl[ \hat{T}^{\rm gf}_{00} \Bigr]_g
	={}&
	\Bigl[ \hat{T}_{00}^{\rm gf}  \Bigr]_{\scr \rm W}
	+ 
	a^{2-D} \, {\rm Re} \Bigl(
		\bigl[ \hat{K}_2 , \hat{A}_0 \bigr]
		+ \bigl[ \partial_k \hat{K}_1 , \hat{A}_k \bigr]
		\Bigr)
		\, ,
\\
\Bigl[ \hat{T}^{\rm gf}_{0i} \Bigr]_g
	={}&
	\Bigl[ \hat{T}_{0i}^{\rm gf}  \Bigr]_{\scr \rm W}
	+ a^{2-D} \, {\rm Re} \Bigl(
		\bigl[ \partial_{i} \hat{K}_1 , \hat{A}_{0} \bigr]
		+ \bigl[ \hat{K}_2 , \hat{A}_{i} \bigr]
	\Bigr)
		\, ,
\\
\Bigl[ \hat{T}^{\rm gf}_{ij} \Bigr]_g
	={}&
	\Bigl[ \hat{T}_{ij}^{\rm gf}  \Bigr]_{\scr \rm W}
	+ a^{2-D} \,
	{\rm Re}
	\Bigl(
	2 \bigl[ \partial_{(i} \hat{K}_1 , \hat{A}_{j)} \bigr]
	+ \delta_{ij} \bigl[ \hat{K}_2 , \hat{A}_0 \bigr]
	- \delta_{ij} \bigl[ \partial_k \hat{K}_1 , \hat{A}_k \bigr]
		\Bigr)
		\, .
\end{align}
\label{g-W difference}%
\end{subequations}
The Weyl-ordered parts above can easily be defined from the covariant 
expression~(\ref{Tgf def}) directly, by simply symmetrizing the products
of Hermitian operators,
\begin{align}
\Bigl[ T_{\mu\nu}^{\rm gf} \Bigr]_{\scr \rm W}
	={}& 
	- \frac{ 1 }{\xi} 
		\Bigl\{ \hat{A}_{(\mu} ,
		\Bigl( \nabla_{\nu)} + 2 \zeta n_{\nu)} \Bigr)
		\Bigl( \nabla^\rho \hat{A}_\rho - 2 \zeta n^\rho \hat{A}_\rho \Bigr) \Bigr\}
\label{T Weyl}
\\
&	\hspace{0.2cm}
	+
	\frac{ g_{\mu\nu} }{2\xi} 
		\biggl[
		\Bigl\{ \hat{A}_\rho ,  \Bigl( \nabla^\rho + 2 \zeta n^\rho \Bigr)
		\Bigl( \nabla^\sigma \hat{A}_\sigma - 2 \zeta n^\sigma \hat{A}_\sigma \Bigr) \Bigr\}
	+
	\Bigl( \nabla^\rho \hat{A}_\rho
		- 2 \zeta n^\rho \hat{A}_\rho \Bigr)^{\!2} \,
	\biggr]
	\, .
\nonumber 
\end{align}
The nonvanishing commutators in~(\ref{g-W difference})
that are just~$c$-numbers, can be evaluated as coincidence
limits of position space commutators using~(\ref{subs 1})--(\ref{single condition 2}). This leads 
to the following relation between the properly gauge-ordered gauge-fixing contribution to
the energy-momentum tensor and the Weyl-ordered contribution,
\begin{align}
\Bigl[ \hat{T}_{\mu\nu}^{\rm gf} \Bigr]_g
	={}& \Bigl[ \hat{T}_{\mu\nu}^{\rm gf} \Bigr]_{\scr \rm W}
	- 
	{\rm Re}
	\Biggl\{
	\biggl[
	2 \Bigl( \nabla_{(\mu} \nabla'_{\nu)} + 2 \zeta n_{(\mu} \nabla'_{\nu)} \Bigr)
\nonumber \\
&	\hspace{2.5cm}
	- g_{\mu\nu} \Bigl( \nabla^\rho \nabla'_\rho + 2 \zeta n^\rho \nabla'_\rho \Bigr)
	\biggr]
	\biggl[
	\Bigl( \frac{a'}{a} \Bigr)^{\!\zeta} i \bigl[ \tensor*[^{\scr \! - \!}]{\Delta}{^{\scr \! +\! }} \bigr](x;x')
	\biggr]_{x'\to x}
	\Biggr\}
	\, .
\label{Tgf Weyl}
\end{align}
The result in~(\ref{Tgf vanishing}) is guaranteed only if we take the left-hand side
of the expression above as the definition, or if we take the full right-hand side.
This shows how the Weyl-ordered product is incomplete when defining the observable,
as it needs to be supplemented by the contribution in the second line of~(\ref{Tgf Weyl}).
This $c$-number contribution can, in fact, be recognized as the FP ghost contribution
to the energy-momentum tensor that descends from the FP ghost 
action~(\ref{ghost action}).
Thus, we see that even in the linear Abelian gauge theory,
and even without being explicitly introduced, the FP ghosts naturally arise
as commutators accounting for the difference between proper operator ordering of operators and Weyl
ordering of operators. Either form of the definition we adopt,~(\ref{Tmunu operator})
or~(\ref{Tgf Weyl}), leads to the vanishing expectation value~(\ref{Tgf vanishing}).
Therefore, one has two options: either gauge-order operators containing constraints,
or Weyl-order all the products~{\it and} add the FP ghost contributions.

Had we ignored the question of operator ordering and defined~(\ref{T Weyl}) as the observable,
its expectation value would no longer vanish identically. This is what is behind the conclusion
in~\cite{BeltranJimenez:2008enx,BeltranJimenez:2009oao,BeltranJimenez:2011nm}
that the gauge-fixing contribution to the energy-momentum tensor of the photon engenders
a nonvanishing cosmological constant contribution. This conclusion, however, does not hold up. 
Reference~\cite{Zhang:2022csl} attempted to address the question of the vanishing gauge-fixing 
contribution to the energy-momentum tensor
in de Sitter space
by considering Weyl ordering of operators, but without introducing FP ghosts
and instead employing adiabatic subtraction to 
obtain~(\ref{Tgf vanishing}). That approach cannot be correct, as it suggests that
gauge-independence has something to do with the divergent UV structure of the theory,
and it leaves the option 
for the gauge-fixing part of the energy-momentum tensor to produce a physical 
contribution
in some spacetime other than de Sitter. Moreover, their conclusion that FP ghosts are not necessary
when operators are Weyl ordered contradicts the results of this section, as well as
contradicting consistent 
analogous for Stueckelberg vector fields~\cite{Belokogne:2015etf,Belokogne:2016dvd}.

\section{Discussion}
\label{sec: Discussion}

In this work we considered the canonical quantization of the photon (massless vector field)
in spatially flat FLRW spacetimes
in the two-parameter family of linear gauges~(\ref{gauge-fixing term}).
We used this framework to demonstrate that observables with appropriate operator ordering 
respect expectations set by the correspondence principle. In particular, we had 
demonstrated how the gauge-fixing term~(\ref{gauge-fixing term}) does not
contribute to the energy-momentum tensor expectation value for any physical state.

Along the way we had elucidated how the canonical formulation is the appropriate framework
for the Gupta-Bleuler quantization, and had derived the subsidiary condition on the
physical state from the first-class constraint structure of the classical theory.
This subsidiary condition translates into subsidiary conditions on two-point functions
derived in Sec.~\ref{subsec: General properties}. The form of these subsidiary conditions 
depends on the two gauge-fixing parameters~$\xi$ and~$\zeta$ from~(\ref{gauge-fixing term}),
as in fact does the two-point function itself. Two-point functions with free gauge-fixing parameters
are particularly useful for computations, as they allow for explicit checks of gauge independence
of observables, that should not depend on the gauge-fixing parameters. The construction of 
inflationary gauge-independent quantum observables in still in its early stages~\cite{Miao:2017feh}.

Different two-point functions considered in Sec.~\ref{sec: Two-point functions} are basic building blocks 
of nonequilibrium loop computations in the Schwinger-Keldysh formalism appropriate for
early universe cosmology. We believe that the framework set up in this paper
will facilitate the construction of photon propagators in realistic inflationary spacetimes,
that will in turn allow the investigation of slow-roll corrections to large infrared effects found
when photons interact with spectator scalars and gravitons in de Sitter~\cite{Prokopec:2002jn,Prokopec:2002uw,Prokopec:2003bx,Prokopec:2003iu,Kahya:2005kj,Kahya:2006ui,Prokopec:2006ue,Prokopec:2007ak,Prokopec:2008gw,Leonard:2012si,Leonard:2012ex,Chen:2016nrs,Chen:2016uwp,Chen:2016hrz,Kaya:2018qbj,Popov:2017xut,Glavan:2019uni,Leonard:2013xsa,Glavan:2013jca,Wang:2014tza,Glavan:2015ura,Glavan:2016bvp,Miao:2018bol}. We had expressed the photon two-point function in 
Sec.~\ref{subsec: Mode sum representation} in terms of several scalar mode functions.
The equations of motion that these scalar mode functions satisfy are collected in Sec.~\ref{subsec: Solving for dynamics}.
In that form it should be considerably easier to identify convenient choices for gauge-fixing parameters
that lead to simpler propagators for given inflationary backgrounds. Constructing such 
two-point functions will be the subject of future work~\cite{Domazet:2023}. Furthermore, 
collecting subsidiary conditions for two-point functions in Sec.~\ref{subsec: General properties}, 
in both the double derivative and the single derivative forms, 
will facilitate consistency checks
of photon two-point functions constructed in future studies.

Before embarking to compute the photon propagators in power-law and slow-roll inflation
it is instructive to check  the existing two-point functions in de Sitter space ($\epsilon\!=\!0$) from the literature
versus subsidiary conditions for the Wightman function~(\ref{all double conditions W}) 
and for the Feynman propagator~(\ref{all double conditions F}). 
The checks are summarized in Table~\ref{consistency conditions}.
\begin{table}[h!]
\renewcommand{\arraystretch}{1.3}
\centering
\begin{tabular}{l  c  c  c  }
\hline
\quad
de Sitter propagator 
	&
	$\ \ \bigl\langle \hat{\Pi}_0 \hat{\Pi}_0 \bigr\rangle \ \ $
	&
	$\ \bigl\langle \hat{\Pi}_0 \partial_i \hat{\Pi}_i \bigr\rangle \ $
	&
	$\bigl\langle \partial_i \hat{\Pi}_i \partial_j \hat{\Pi}_j \bigr\rangle $
\\
\hline
\hline
\quad
Allen-Jacobson~\cite{Allen:1985wd} ($\xi\!=\!1, \zeta\!=\!0,D$) $\quad$
	&
	$\boldsymbol{\times}$
	&
	$\checkmark$
	&
	$\checkmark$
\\
\hline
\quad
Tsamis-Woodard~\cite{Tsamis:2006gj} ($\xi\!=\!0, \zeta\!=\!0,D$)
	&
	$\checkmark$
	&
	$\checkmark$
	&
	$\checkmark$
\\
\hline
\quad
Youssef~\cite{Youssef:2010dw} ($\xi,\zeta\!=\!0,D\!=\!4$)
	&
	$\boldsymbol{\times}$
	&
	$\checkmark$
	&
	$\checkmark$
\\
\hline
\quad
Fr\"ob-Higuchi~\cite{Frob:2013qsa} ($\xi, \zeta\!=\!0, D$)
	&
	$\boldsymbol{\times}$
	&
	$\checkmark$
	&
	$\checkmark$
\\
\hline
\quad
Woodard~\cite{Woodard:2004ut} ($\xi\!=\!1, \zeta\!=\!1, D$)
	&
	$\checkmark$
	&
	$\checkmark$
	&
	$\checkmark$
\\
\hline
\end{tabular}
\caption{
Checks of subsidiary conditions for massless photon two-point functions in de 
Sitter reported in the literature. The checks refer to both the Wightman functions
and Feynman propagators. The first four entries with~$\zeta\!=\!0$ correspond to covariant 
gauges, while the fifth entry with~$\zeta\!=\!1$ corresponds to a conformal gauge in~$D\!=\!4$.
 }
\label{consistency conditions}
\end{table}
Unexpectedly, they reveal that the general
covariant gauge two-point functions satisfy all the required subsidiary conditions only in the 
exact transverse gauge limit~$\xi\!\to\!0$. For nonvanishing~$\xi$ the double divergence
of the covariant gauge two-point functions fails to vanish off-coincidence,
\begin{equation}
\nabla^\mu \nabla'^\nu
	i \bigl[ \tensor*[_\mu^{\scr - \! }]{\Delta}{_\nu^{\scr \! +}} \bigr] (x;x') 
	=
	- \frac{\xi H^D \, \Gamma(D) }{ (4\pi)^{\frac{D}{2}} \Gamma\bigl( \frac{D}{2} \bigr) } 
	=
\nabla^\mu \nabla'^\nu
	i \bigl[ \tensor*[_\mu^{\scr + \! }]{\Delta}{_\nu^{\scr \! +}} \bigr] (x;x') \, ,
	\qquad
	x \neq x' \, ,
\end{equation}
violating subsidiary conditions~(\ref{double condition 1}) and~(\ref{double condition F 1}).
This points to inconsistencies of known results even for the relatively simple case of
the maximally symmetric de Sitter space. These inconsistencies will be examined in more
detail and addressed elsewhere~\cite{Glavan:2022dwb,Glavan:2022nrd}.

\section*{Acknowledgments}

I am grateful to Chun-Yen Lin, Danijel Jurman, and Jose Beltr\'{a}n Jim\'{e}nez for 
discussions on the topic, and especially to Tomislav Prokopec for the critical reading of the manuscript.
This work was supported by the Czech Science 
Foundation (GA\v{C}R) Grant No.~20-28525S.


\end{document}